\documentclass[manuscript]{acmart}
\AtBeginDocument{%
  }

\setcopyright{acmlicensed}
\copyrightyear{2025}
\acmYear{2025}
\acmDOI{XXXXXXX.XXXXXXX}
\acmConference[CHI'26]{The ACM CHI conference on Human Factors in Computing Systems}{13–17 April, 2026}{Barcelona}
\acmISBN{978-1-4503-XXXX-X/2018/06}

\usepackage{booktabs}  
\usepackage{colortbl}  
\usepackage{xcolor}    
\usepackage{subcaption}
\usepackage{svg}
\usepackage{graphicx}
\usepackage{tikz}
\usetikzlibrary{arrows.meta,positioning,fit,calc,quotes}
\usepackage{tikz}
\usetikzlibrary{positioning,calc}

\definecolor{pretestBlue}{RGB}{123,160,214}
\definecolor{trainPurple}{RGB}{170,155,197}
\definecolor{postPeach}{RGB}{229,196,156}

\tikzset{
  labelL/.style = {anchor=west, font=\large},
  bullet/.style = {anchor=west, font=\large},
  small/.style  = {font=\normalsize},
  box/.style    = {draw=black, rounded corners=3pt, minimum width=1.5cm,
                   minimum height=0.8cm, inner sep=2pt},
  intro/.style  = {box, fill=white},
  pre/.style    = {box, fill=pretestBlue!80},
  train/.style  = {box, fill=trainPurple!70},
  test/.style   = {box, fill=trainPurple!90},
  post/.style   = {box, fill=postPeach!80},
  leveltag/.style = {anchor=east, font=\large, gray},
}

\usepackage{graphicx}

\definecolor{LPline}{HTML}{A23B72}   
\definecolor{LPfill}{HTML}{F6EAF1}   
\definecolor{HPline}{HTML}{2E86AB}   
\definecolor{HPfill}{HTML}{E8F1F7}   

\tikzset{
  >=Latex,
  panel/.style={draw=black, rounded corners=2pt, inner sep=6pt},
  title/.style={font=\bfseries\small},
  percent/.style={font=\scriptsize, fill=white, inner sep=1pt},
  stateLP/.style={rounded corners=6pt, minimum width=30mm, minimum height=9mm,
                  draw=LPline, fill=LPfill, line width=0.9pt, align=center,
                  font=\scriptsize\bfseries},
  stateHP/.style={rounded corners=6pt, minimum width=30mm, minimum height=9mm,
                  draw=HPline, fill=HPfill, line width=0.9pt, align=center,
                  font=\scriptsize\bfseries},
  edgeLP/.style={->, very thick, draw=LPline},
  edgeHP/.style={->, very thick, draw=HPline},
  looplab/.style={font=\scriptsize}
}

\usepackage{caption}

\usepackage{float}
\usepackage{multirow}
\usepackage[table]{xcolor}
\definecolor{gray2}{rgb}{0.9, 0.9, 0.9}
\usepackage{booktabs}
\usepackage[table]{xcolor}
\usepackage{makecell}
\usepackage{multirow}
\renewcommand{\arraystretch}{1.2}
\newcommand{\armhdr}[3]{\makecell[c]{\textbf{#1}\\ \textbf{\(N=\) #3}}}

\begin{document}

\title{Exploring the Design and Impact of Interactive Worked Examples for Learners with Varying Prior Knowledge}

\author{Sutapa Dey Tithi}
\email{stithi@ncsu.edu}
\orcid{0009-0007-5815-882X}

\affiliation{
  \institution{North Carolina State University}
  \city{Raleigh}
  \state{NC}
  \country{USA}
}
\author{Xiaoyi Tian}
\email{xtian9@ncsu.edu}
\orcid{0000-0002-5045-0136}

\affiliation{
  \institution{North Carolina State University}
  \city{Raleigh}
  \state{NC}
  \country{USA}
}
\author{Ally Limke}
\email{anlimke@ncsu.edu}

\affiliation{
  \institution{North Carolina State University}
  \city{Raleigh}
  \state{NC}
  \country{USA}
}
\author{Min Chi}
\email{mchi@ncsu.edu}

\affiliation{
  \institution{North Carolina State University}
  \city{Raleigh}
  \state{NC}
  \country{USA}
}

\author{Tiffany Barnes}
\email{tmbarnes@ncsu.edu}

\affiliation{
  \institution{North Carolina State University}
  \city{Raleigh}
  \state{NC}
  \country{USA}
}
\renewcommand{\shortauthors}{Tithi et al.}

\begin{abstract}
Tutoring systems improve learning through tailored interventions, such as worked examples, but often suffer from the aptitude-treatment interaction effect where low prior knowledge learners benefit more. We applied the ICAP learning theory to design two new types of worked examples, Buggy (students fix bugs), and Guided (students complete missing rules), requiring varying levels of cognitive engagement, and investigated their impact on learning in a controlled experiment with 155 undergraduate students in a logic problem solving tutor. Students in the Buggy and Guided examples groups performed significantly better on the posttest than those receiving passive worked examples. Buggy problems helped high prior knowledge learners whereas Guided problems helped low prior knowledge learners. Behavior analysis showed that Buggy produced more exploration-revision cycles, while Guided led to more help-seeking and fewer errors. This research contributes to the design of interventions \textcolor{black}{in logic problem solving} for varied levels of learner knowledge and a novel application of behavior analysis to compare \textcolor{black}{learner interactions with the tutor}.
\end{abstract}


\begin{teaserfigure}
  \includegraphics[width=\textwidth]{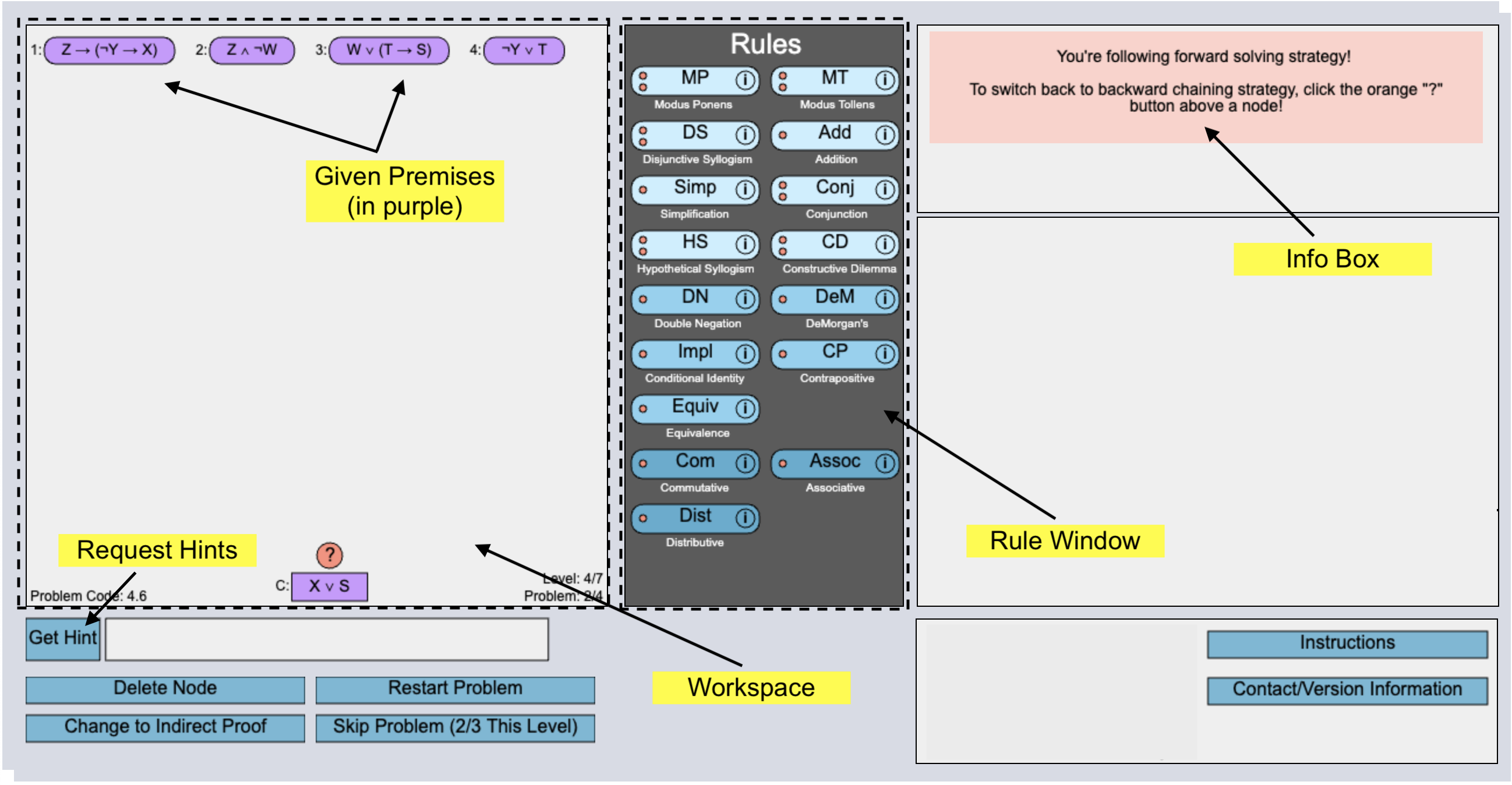}
  \caption{The tutor interface with the integrated three-panel design: student workspace (left), domain rules (center), and contextual information/instructions panel (right). The hint system and feedback message appear in the bottom-left corner.}
  \Description{The tutor interface with the integrated three-panel design: student workspace (left), domain rules (center), and contextual information/instructions panel (right). The hint system and feedback message appear in the bottom-left corner.}
  \label{fig1}
\end{teaserfigure}



\maketitle

\section{Introduction}

Intelligent tutoring systems (ITSs) provide adaptive scaffolding and personalized interventions, demonstrating significant positive impacts on learning outcomes \cite{murray2003overview, ma2014intelligent}. However, the effectiveness of instructional interventions is not uniform across all learners. Research has found evidence of aptitude-treatment interaction (ATI) \cite{cronbach1981aptitudes,snow1991aptitude}, meaning that the effectiveness of specific instructional strategies varies significantly based on individual learner characteristics. These aptitude treatment interactions manifest across multiple dimensions of learner differences, including prior experience level, prior working memory, and incoming self-regulated learning ability \cite{kalyuga2001problem,lehmann2016working,fuchs2019using,yeh2015aptitude}.

ITSs often incorporate two primary instructional approaches: Problem Solving (PS), where students independently construct solutions, or Worked Examples (WE), where the tutor presents step-by-step solutions for students to study. With regard to learners’ abilities, research within ITSs has shown that high-ability learners can benefit from less guidance or more independent problem-solving practice. In contrast, low-ability learners benefit from more specific and direct guidance \cite{arroyo2000macroadapting,luckin1999ecolab}. Worked examples reduce the intrinsic cognitive load on learners and improve their learning \cite{paas2003cognitive}. Moreover, examples can save learners time without reducing their learning \cite{mostafavi2015data}. However, they may not be beneficial or could even be detrimental for learners with high prior knowledge, a phenomenon known as the expertise reversal effect \cite{nievelstein2013worked,kalyuga2009expertise}. Thus, designing instructional interventions faces the challenge of creating opportunities so that learners of any incoming proficiency can master the same material and improve their learning outcomes in a similar amount of time. 

The ICAP (Interactive, Constructive, Active, Passive) framework provides a theoretical lens for addressing this challenge \cite{chi2014icap}. This theory links students' cognitive engagement to their learning outcomes. It states that learning outcomes systematically increase as cognitive engagement increases, from \textit{Passive} (receiving information), to \textit{Active} (manipulating materials), to \textit{Constructive} (generative), to \textit{Interactive} (co-generative). Traditional worked examples, requiring only passive observation, fall into the lowest level of this hierarchy. There remain critical gaps in understanding: (1) how to design worked examples that promote higher levels of cognitive engagement while maintaining their scaffolding benefits, (2) empirical evidence of how worked examples with different levels of cognitive engagement help learners. 

To address these gaps, we designed and empirically evaluated two types of interactive worked examples within an intelligent logic tutoring system: Buggy (students identify and fix bugs in worked solutions, promoting active engagement through debugging and fixing), and Guided (students reconstruct missing parts in worked solutions with step-by-step hints, promoting active reconstruction of proofs). The tutor also provides a baseline \emph{Passive} WE and a \emph{Constructive} PS intervention. \textcolor{black}{The student-tutor interaction design and desired cognitive engagement in Buggy and Guided examples} follow the ICAP framework and aim to increase interaction with the tutor relative to passive WE while maintaining more structure than independent PS. \textcolor{black}{Propositional logic proofs are open-ended with multiple valid solutions. Our Buggy examples present expert-designed erroneous solutions without marking the incorrect steps, requiring students to understand both local rule applications and global proof coherence to successfully debug and fix incorrect steps. Guided examples decompose the solution into manageable chunks or subgoals, providing step-specific hints to help reconstruct the proof. Both Buggy and Guided examples are \textit{Interactive} WEs, since they provide immediate feedback on student work. These two interventions are novel in that intelligent problem solving environments are often custom-built to enable the tracking of student work, and rarely have the capability to provide erroneous examples or subgoal-focused, partially-worked guided examples.}

We integrated these interactive worked examples (Buggy, Guided) in an intelligent logic tutor along with passive Worked Example (WE) and constructive Problem Solving (PS). We investigated the impact of these worked examples on students' learning \textcolor{black}{of propositional logic problem solving} in a controlled experiment with 155 undergraduate Computer Science (CS) students. Students in different conditions practiced interleaving PS with one of: WE, Buggy, or Guided examples. We recorded students' clickstream interactions in the tutor and analyzed the log data. We investigated students' performance in posttest problems after training with three different interventions and analyzed further to answer the following research questions:

\begin{itemize}
    \item RQ1 (Learning Outcomes): How do worked examples targeting different engagement levels (i.e., passive WE, Buggy examples, and Guided examples) affect students' performance in posttest problems?
    \item RQ2 (Moderation by Prior Knowledge): How do these effects vary with students' prior knowledge?
    \item RQ3 (Problem Solving Behavior): How do these interventions impact students' problem solving behaviors during training?
\end{itemize}

The contributions of this work are: (1) two theory-driven novel interactive worked example interface scaffolding designs \textcolor{black}{in propositional logic learning} promoting higher cognitive engagement; (2) empirical evidence of differential effectiveness of worked example interventions based on prior knowledge, informing adaptive \textcolor{black}{logic} tutoring; (3) a Markov model based visualization to interpret problem-solving behavior patterns; (4) design implications for worked examples that support guided proof construction, evaluation and debugging practice. 

\section{Related Work}
\subsection{Instructional Support in Intelligent Tutoring Systems}
Multiple approaches have been explored within intelligent tutoring systems (ITSs) to facilitate student learning \cite{singh2019implementation}. These include Problem Solving (PS), where students solve the problem themselves, and Worked Example (WE), where the problem is solved for them step-by-step. WEs have been shown to reduce intrinsic cognitive load and improve learning \cite{paas2003cognitive}. However, the research has shown that WEs have mixed efficacy; while they may not be beneficial to students with high prior knowledge when problems are structured \cite{nievelstein2013worked}, they have also been shown to be most effective when they provide instructional explanations or rationales for the solution steps \cite{renkl2005worked}.

This phenomenon can be explained by Aptitude-Treatment Interaction (ATI) theory, which states that optimal learning occurs when instructional approaches align with individual learner characteristics \cite{cronbach1981aptitudes}. Aptitudes can be broadly defined as any individual difference that affects learning, and these aptitudes interact with instructional approaches in complex, often non-linear ways \cite{snow1991aptitude}. Interventions that benefit one type of learner may actually harm another's performance. Although instructional methods such as WEs benefit novices, they may not be helpful to students with higher expertise. As students develop expertise, they build mental schemas that allow them efficient information processing. When experts receive the same detailed guidance that helps novices, they must process information they already know, wasting cognitive resources and thus reducing performance \cite{kalyuga2009expertise,oksa2010expertise}.

The Cognitive Load theory identifies three types of cognitive load: intrinsic load (inherent to the material and learner's expertise level); extraneous load (imposed by poor design or irrelevant tasks); and germane load (productive effort supporting schema construction) \cite{sweller2011cognitive}. WEs are designed to reduce intrinsic cognitive load and improve learning efficiency for novices \cite{sweller1988cognitive,kalyuga2009expertise}. As learners progress towards higher expertise, techniques like requiring self-explanations help transition from example study to independent problem solving \cite{renkl2005worked,chi1994eliciting}, and can help achieve lower order learning objectives (remember/understand/early apply) \cite{bloom2010taxonomy}. For higher order objectives (analyze/evaluate/create), more active problem solving and structured interventions are typically more effective than fully WEs \cite{chi2014icap}.

\subsection{ICAP Learning Framework}
The ICAP learning framework differentiates cognitive engagement activities into four modes: Interactive, Constructive, Active, and Passive \cite{chi2014icap}. Passive engagement includes activities where learners only receive information from the instructional materials, such as reading texts or watching instructional videos. Active engagement includes activities where students manipulate some parts of the learning materials, by copying some of the problem solution steps \cite{vanlehn2007tutorial}, or choosing a justification from a menu of options \cite{conati2000toward}, and so forth. Constructive engagement requires learners to generate additional externalized outputs beyond what was provided in the learning materials, such as, comparing and contrasting cases \cite{schwartz1998time}, asking questions \cite{graesser1994question}, etc. Finally, interactive engagement involves collaborative knowledge construction such as explaining to each other through dialogue with peers, tutors, or human-AI collaboration with intelligent systems \cite{roscoe2007understanding}. The framework states that as students become more engaged with the learning materials, from passive to active to constructive to interactive, their learning will increase. Chi and Wylie presented empirical evidence from multiple studies to validate their ICAP hypothesis \cite{chi2014icap}. Their first study examined all four engagement modes within materials science instruction, demonstrating improved learning of 8-10\% with each progressive mode. Then they presented two other studies that implemented passive, active, and constructive modes in the domains of evolutionary biology and plate tectonics, finding similar patterns of enhanced learning with increased engagement. Finally, they presented another study with pairwise comparisons across note-taking, concept mapping, and self-explanation yielding similar observations. 

The default training intervention in our intelligent tutor allows students to practice propositional logic proofs independently with PS or traditional WEs. Similarly to solving programming problems, solving logic proofs requires students to understand a system of domain principles or rules and to apply them in a sequence to achieve a goal. In our tutor, when students solve WEs, the tutor shows every step with explanations, and students click through to progress forward or revisit the previous steps. In PS, on the other hand, students themselves need to apply appropriate domain rules to construct new steps towards achieving the goal. In our tutor, we integrated two new types of worked examples: Buggy (students fix bugs in worked solutions), and Guided (students reconstruct missing parts in worked solutions presented in Parsons problem fashion \cite{ericson2022parsons}). Both mechanisms are designed to deepen engagement and potentially improve learning outcomes. We elaborate on our design of Buggy and Guided examples in Section \ref{systemdesign}.

In addition to traditional PS and WE interventions, prior studies explored the effectiveness of erroneous examples and Parsons problems\textcolor{black}{, particularly in domains such as programming and arithmetic}. Buggy or erroneous examples leverage common mistakes made by students that can promote deeper understanding \cite{grosse2007finding, richey2019more}. When augmented with incorrect examples, algebra tutors enhanced student learning compared to those with only correct examples \cite{booth2013using}. Similarly, studies with the AdaptErrEx decimals tutor showed enhanced learning when incorrect examples targeted common misconceptions \cite{adams2014using,mclaren2015delayed}. However, effectiveness may depend heavily on prior knowledge\textemdash novices may lack the necessary schemas to identify errors, while advanced learners may benefit from the critical thinking required for debugging. Well-designed and higher-level feedback, instead of giving away the solution, can facilitate productive debugging behavior and help students understand their problems \cite{suzuki2017exploring}. Moreover, visualization has been found to be a good means for novice program debugging \cite{brusilovsky1993program}. \textcolor{black}{ However, most existing Buggy example research focuses on short, single-step skills (e.g., arithmetic, decimals) or debugging in programming, and very few studies examine Buggy interventions in complex, multi-step problem solving tasks such as propositional logic proof construction.} In a Parsons problem, students rearrange jumbled solution steps or complete the missing parts \cite{ericson2022parsons}. Research in programming education shows that Parsons problems (rearranging and completing partially blank lines of code into a valid program) are an effective exercise interface for teaching programming patterns, significantly improving overall code-writing abilities \cite{weinman2021improving,karavirta2012mobile}. Recent work has extended this approach to mathematical proofs, demonstrating reduced difficulty in proof construction \cite{poulsen2022evaluating}. Understanding high-level contextual significance \cite{prather2022scaffolding} and subgoal labels \cite{morrison2016subgoals} can help students solve Parsons problems and improve their learning outcomes. Data-driven, subgoal-oriented Parsons problems can enhance students' problem decomposition-recomposition skills in solving propositional logic proofs \cite{shabrina2023learning}. However, students struggle with Parsons problems when they first encounter this type of structured or chunked problem or when the connections among different parts of the problem are complex \cite{shabrina2023learning}. \textcolor{black}{ While worked and partially worked examples exist in prior ITS research, their adaptation to other complex problem solving areas remains underexplored. Our Guided examples specifically target problem decomposition-recomposition skills while providing step-specific hints, reducing cognitive load in a complex problem solving environment.}

In this study, we investigate the design and the impact of Buggy examples and Guided examples along with traditional WE and PS in our tutor. \textcolor{black}{Our contribution is a systematic ICAP-aligned design of three worked-example types—passive, active, and constructive using identical learning materials for each representation, enabling a controlled comparison of engagement modes.} However, identifying the mode of a particular activity can be a non-trivial task \cite{gaweda2021student}. Our traditional WEs help students learn new information with \textit{passive} engagement, and our PS helps students practice logic proofs through independent proof \textit{construction}. In our design, Guided examples fade away or remove some connections in worked solutions, and students apply domain principles to complete missing connections. This task requires them to have an understanding of the given solution as a whole to deduce the domain principles that need to be applied to the missing part of the solution. Thus, we classify our Guided examples as an \textit{active} ICAP mode. On the contrary, Buggy examples present worked solutions with expert-designed bugs, and students identify the bugs and fix them by typing the correct element. \textcolor{black}{ Unlike prior buggy example implementations, our design targets multi-step propositional logic proof construction, where debugging requires understanding of both local rule applications and global proof coherence.} The task of finding bugs requires students to assess the given solution and produce qualifications such as ``correct'' or ``incorrect'', and the task of correcting the bugs needs them to add new information to the solution; thus, we classify these tasks as \textit{constructive} ICAP mode. However, as we mentioned previously, classifying the mode of a particular activity is not trivial; Guided or Buggy examples can also be considered \textit{interactive} depending on the context, since students often rely on the tutor's feedback or hints during the proof construction or the debugging process.

The increasing capabilities of generative AI systems create additional challenges and opportunities in education. Classroom and lab studies show both benefits and challenges: while generative AI-powered assistants can provide conceptual scaffolding without revealing complete solutions \cite{kazemitabaar2024codeaid}, novices often struggle describing their intent in prompts, evaluating the correctness of generated solutions, and editing prompts when the generated code is incorrect \cite{nguyen2024beginning}. Recent studies also examined how LLM-generated programming hints of varying granularity support or disappoint novices \cite{xiao2024exploring}. While human-AI collaboration with generative AI systems can help students in their learning process \cite{pesovski2024generative}, LLMs are prone to hallucinating content that can be misaligned with course materials \cite{jia2024assessing}. Thus, students need skills in evaluating, correcting, and optimizing any given content during human-AI collaborative problem solving. This creates new demands for educational systems that develop higher-order thinking skills rather than just problem-solving from scratch. \textcolor{black}{ Logic problem solving is a multi-step and open-ended procedure. In this work, our interaction designs target} enhancing students' logic proof evaluation, debugging, and reconstruction capabilities while respecting individual differences in learning abilities.

\section{System Design} 
We designed two different instructional problem solving methods following the ICAP learning framework: Buggy examples (Buggy) and Guided examples (Guided). All methods use the same underlying propositional logic problems and require students to engage with the same set of domain principles/rules. 

\subsection{Tutor Overview}
Our tutor is used in the context of an undergraduate Discrete Mathematics course where students practice open-ended, multi-step propositional logic problems. Students work within a three-panel interface as shown in Figure \ref{fig1}. The left panel provides the primary workspace where students construct proofs by deriving new logical statements, represented as nodes (circles). The center panel contains the domain rule buttons with their descriptions. Students use the rules to justify their derived statements; each node is justified using a rule and 1-2 parent nodes indicated by edges and a rule name label in the workspace. The right panel displays contextual information and instructions.


The proof construction process is presented in a graphical representation where each logical statement appears as a graphical node connected by edges that represent logical dependencies (as shown in Figure \ref{fig:ps}). The given statements appear at the top in purple, while the target conclusion statement appears in purple at the bottom with a question mark, showing that it is not yet justified. Students construct proofs by deriving intermediate statements, creating a connected path from the givens to the conclusion. Each derivation step requires students to specify both the derived statement and the domain rule applied, and the parent statements supporting the derivation. This graphical representation allows students to visualize the overall structure of their proof.

The tutor includes four sections: introduction, pretest, training, and posttest. The introduction consists of two problems where students get acquainted with the interface. Then, the students take the pretest, which consists of two problems. The pretest is used to measure the incoming proficiency of the students and assign them to intervention groups through stratified sampling. The training session then consists of five ordered levels of increasing difficulty, and each level consists of four problems. Finally, the posttest level (level 7) consists of six problems. \textcolor{black}{During the pretest and posttest problems, the tutor does not provide hints for the next step, but offers immediate error messages when students make mistakes while deriving a new step.} For each problem, the students receive a score between 0 and 100 based on efficient proof construction, with higher scores corresponding to attempts with shorter solution size, higher accuracy of rule application, and shorter time.

Based on the intervention, the tutor presents four types of problems during training: Worked Example (WE), Problem Solving (PS), Guided Example (Guided), Buggy Example (Buggy).

\subsection{Design ICAP Interventions} \label{systemdesign}
We integrated new problem types that progress in the ICAP framework, along with our existing problem types, PS and WE. Buggy example (Buggy) is designed to help students learn from active debugging and fixing, and Guided example (Guided) is designed to support active engagement by the construction of missing inference connections. Both problem types also support passive learning as they present an expert solution as a worked-out solution with some perturbations.

\begin{figure}[t]
    \centering
    
    \begin{subfigure}[b]{0.45\textwidth}
        \centering
        \includegraphics[width=\textwidth]{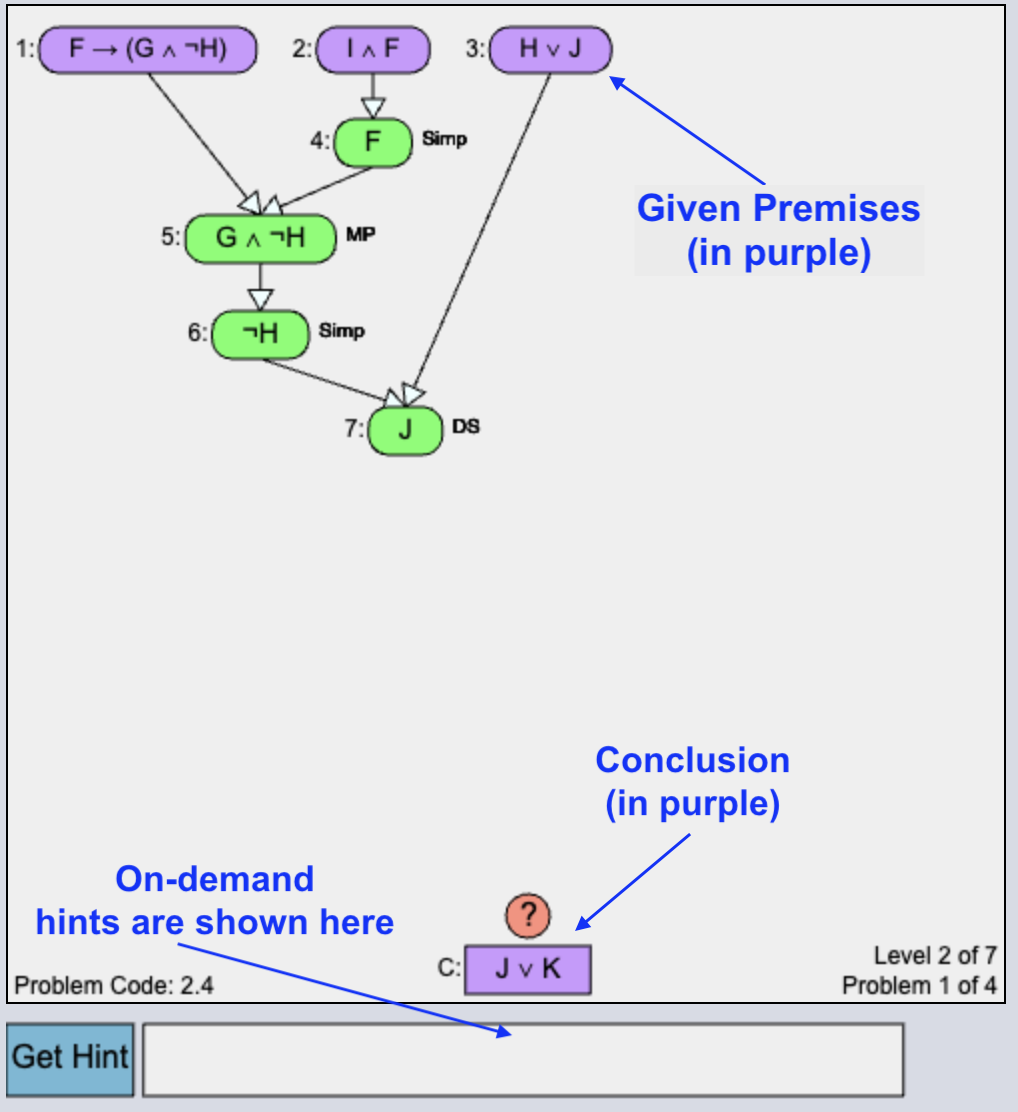}
        \caption{Problem Solving (PS): Students independently derive all proof steps with minimal scaffolding}
        \label{fig:ps}
    \end{subfigure}
    \hspace{0.03\textwidth}
    \begin{subfigure}[b]{0.45\textwidth}
        \centering
        \includegraphics[width=\textwidth]{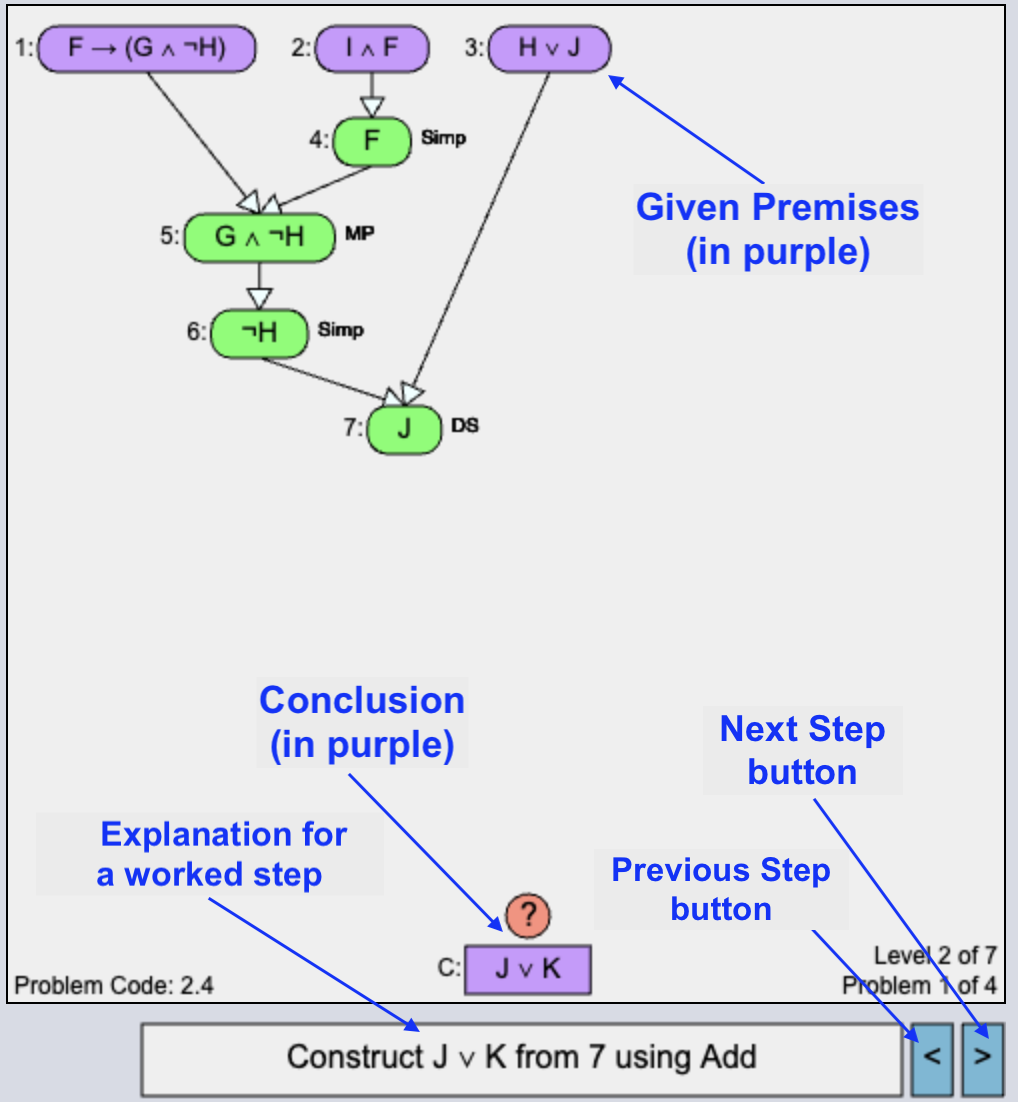}
        \caption{Worked Example (WE): Complete solution demonstrated step-by-step}
        \label{fig:we}
    \end{subfigure}
    
    \vspace{0.02\textheight}
    
    \begin{subfigure}[b]{0.45\textwidth}
        \centering
        \includegraphics[width=\textwidth]{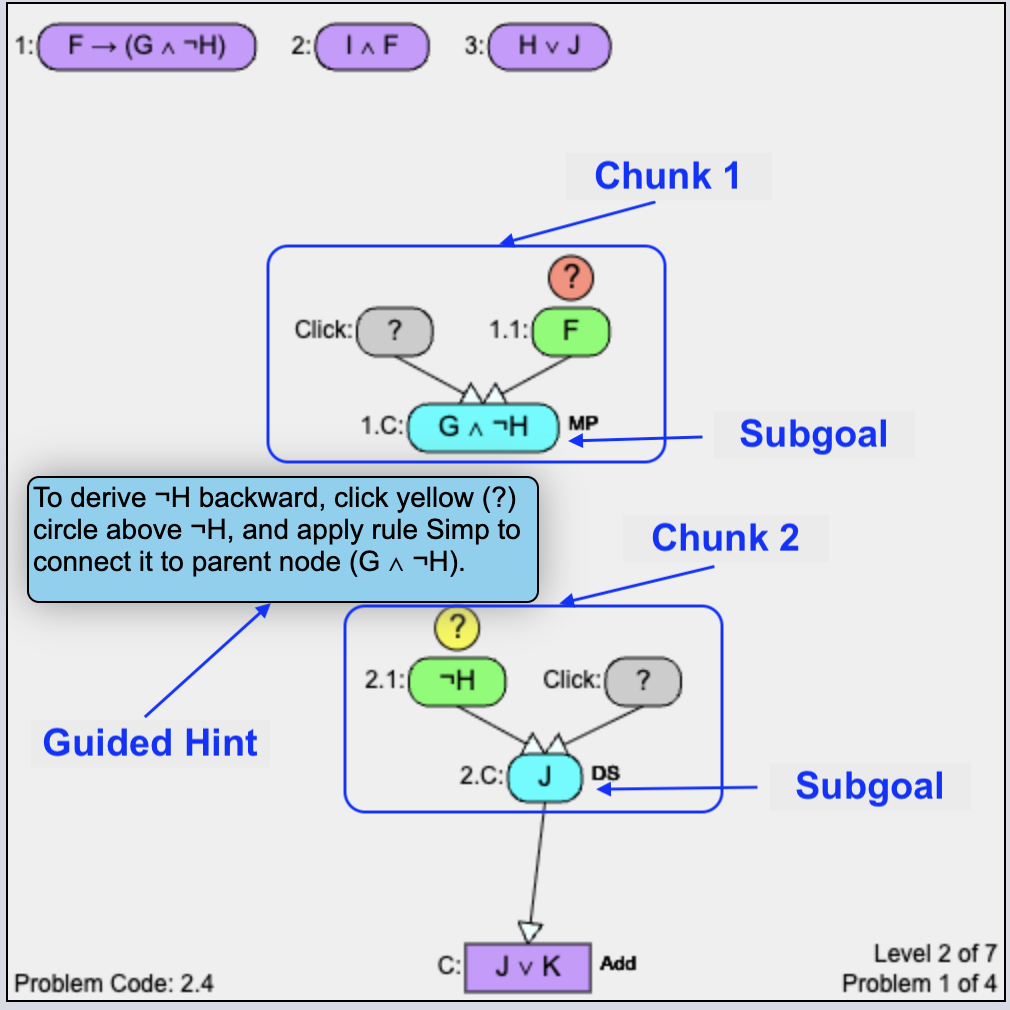}
        \caption{Guided Example (Guided): Partial solution with missing connections}
        \label{fig:gpp}
    \end{subfigure}
    \hspace{0.03\textwidth}
    \begin{subfigure}[b]{0.45\textwidth}
        \centering
        \includegraphics[width=\textwidth]{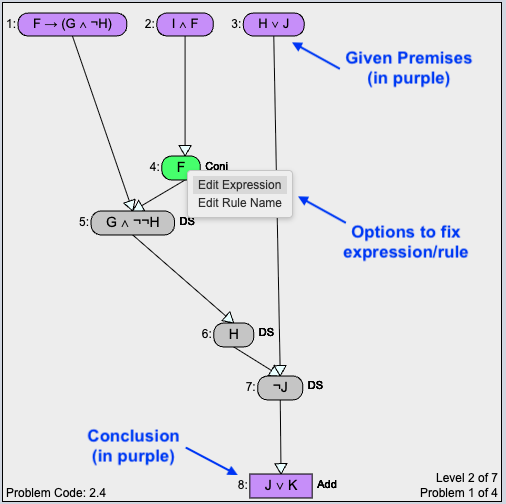}
        \caption{Buggy Example (Buggy): Partial solution with bugs inserted}
        \label{fig:buggy}
    \end{subfigure}

    \caption{The tutor interfaces for four different problem types}
    \label{fig:problem_types}
\end{figure}

\textbf{Problem Solving (PS)} requires students to independently derive all proof steps and thus demands the highest cognitive engagement. Each problem solving step involves the same process: clicking on 1-2 parent nodes and a rule button, and entering the new derived statement, as shown in Figure \ref{fig:ps}. The tutor verifies whether the parent nodes and the selected rule correctly justify the derived statement. Once a step is verified, a new derived node appears in the solution space. The new node is colored based on how often the same node was present in student solutions from previous semesters to this problem, where green means frequent, yellow means infrequent, and gray means never appeared. We call a node ‘necessary’ or ‘needed’ when its deletion would make a solution incomplete. If the tutor detects any incorrect attempt to derive a statement, it provides an error-specific feedback message to the student. PS problem types include a "Get Hint" button, which can be used to request a hint to progress in the solution. A student can get at most four on-demand hints in any training problem.

The design choice of PS reflects the goal of preserving independent problem solving while providing on-demand scaffolding as next-step hints. The frequency-based color coding of the nodes was generated by analyzing student solutions from previous years, and this can help students develop intuitions about common solution approaches without constraining their exploration. 

\textbf{Worked Example (WE)} demonstrates a complete solution, performed by the tutor step by step as the students click on the next step \textbf{(>)} button (Figure \ref{fig:we}). Each step displays the derived statement, the justification rule, the parent statement(s), and indicates a high-level problem-solving operation (construct/extract/transform). The replay functionality allows students to review previous steps with their explanations, addressing a key limitation of traditional worked examples that proceed too quickly for novice understanding.
 
The interface maintains the same visual structure as PS but shifts the student's role from a constructor to an observer. Previously, our tutor supported these two problem types for practicing logic proofs, where PS falls into the `Construction' category and WE falls into the `Passive' category according to the ICAP framework.

\textbf{Designing Buggy and Guided examples.} In Buggy examples, we incorporated errors designed by an expert instructor with 20+ years experience in teaching Discrete Mathematics to change small elements of correct expert solutions. We address these errors as \textit{bugs}. In the correct solutions, the bugs were inserted in two ways: (1) changing the logic operators in a correct statement, (2) changing the rule associated with a statement. The tutor provides contextual instructions about interacting with the interface to fix bugs. Initially, all the statements in the solution are gray, except the givens, which are purple and always correct. As shown in Figure \ref{fig:buggy}, after students click on any node, they can choose from two options: either to fix the statement or the rule associated with the node. Once they select an option, the tutor prompts with a text box, where students can type their answer. After submitting the answer, the tutor verifies its correctness, and identifies the step as correct/incorrect, or the tutor lets them know if that expression/rule was already correct. After the attempt, if the expression/rule is correct, it turns green.

Figure \ref{fig:buggy} shows a screenshot of an intermediate solution state of a Buggy example. The given nodes 1, 2, and 3 are shown at the top of the solution space, and the conclusion node 8 ($J \lor K$) is shown at the bottom. These purple nodes are always correct and cannot be modified. Node 4 ($F$) is green, implying that the student corrected the expression already. Now, if the student again clicks on node 4, the tutor prompts two options; the student may select the option `Edit Rule Name' to replace the given rule (`Conj') with a correct rule (`Simp'). Thus, the student repeats the process of fixing until the solution is entirely correct. We hypothesize that students working with Buggy examples develop the critical evaluation skills of understanding and evaluating the correctness of solutions. This intervention spans through the \textit{passive} (understanding of the given solution), \textit{active} (finding the bugs), and \textit{construct} (typing their own solution elements) learning activities in the ICAP framework. 

In our tutor, Guided example (Guided) demonstrates a partially worked example with step-specific hints attached. In this problem, the proof structure is decomposed into chunks or \textit{subgoals}, and each subgoal chunk groups related statements into logically meaningful units, aligning with Renkl's concept of `meaningful building blocks' \cite{renkl2000studying}. Additionally, we integrate step-specific hints into Guided examples to address the `rationale gap' identified in traditional worked examples \cite{renkl2002worked}. The hints guide students in choosing appropriate domain rules and completing the missing parts of the solution. Thus, Guided examples are designed to maintain low intrinsic load through subgoals and step-specific scaffolding while facilitating active problem solving participation. We hypothesize that students working with Guided examples will better learn the domain principles through an active solution reconstruction process. This intervention spans through the \textit{passive} (understanding of the given solution), \textit{active} (deducing the domain principles to complete the missing parts) learning activities in the ICAP framework.
 
Each Guided example provides students with all the statement nodes needed to complete a proof, but students must add a few justifications to connect all the nodes to one another with missing edges for rules. The nodes without incoming edges are unjustified. Guided examples guide students in justifying each unjustified node by specifying the rule used to derive it. In Figure \ref{fig:gpp}, statement 2.1: $-H$  can be derived from 1.C: $G \land -H$ using rule Simplification. To derive the statement, students are guided with a hint, as shown in the Figure \ref{fig:gpp}. To complete this step, students click on the yellow question mark above 2.1, choose the rule Simplification, and click on statement 1.C to show that there should be an edge from 1.C to 2.1. The solutions are divided into chunks, where important intermediate goals or subgoals in the problem are shown in light blue/cyan, grouped with the nodes used to derive them. For example, in this problem, there are 2 chunks 1 and 2, with subgoals 1.C and 2.C respectively, that are needed to complete the problem. The tutor guides students using popup hints with instructions to work backwards from the conclusion to connect to node 2.C, then connect 2.1 to chunk 1’s conclusion 1.C, and then 1.1 to the givens. 

Table \ref{tab:action-patterns} summarizes the distinct student actions expected in each problem type.

\begin{table}[t]
\centering
\small
\caption{Example Activities in Different Problem Types}
\label{tab:action-patterns}
\renewcommand{\arraystretch}{1.2}
\begin{tabular}{@{}l p{7cm} cccc@{}}
\toprule
\textbf{Student Action} & \textbf{Description} & \textbf{PS} & \textbf{WE} & \textbf{Guided} & \textbf{Buggy} \\
\midrule
\rowcolor{gray!5}
Start Problem & Load problem interface and begin working on the assigned task & \checkmark & \checkmark & \checkmark & \checkmark \\
Read Rule & Hover over rule buttons to view descriptions and understand available logical operations & \checkmark & \checkmark & \checkmark & \checkmark \\
\rowcolor{gray!5}
Hint Request & Click hint button (PS) or hover on the nodes (Guided) to receive tutor help about next steps & \checkmark & \textcolor{gray!40}{---} & \checkmark & \textcolor{gray!40}{---} \\
Correct Step & Correct application of a rule by selecting valid premises, appropriate rule, and correct derived statement & \checkmark & \checkmark & \checkmark & \checkmark \\
\rowcolor{gray!5}
Incorrect Step & Attempt to perform a step with invalid premises, wrong rule, or incorrect derived statement & \checkmark & \checkmark & \checkmark & \checkmark \\
Delete Action & Remove previously entered node to backtrack and explore alternative solution paths & \checkmark & \textcolor{gray!40}{---} & \textcolor{gray!40}{---} & \textcolor{gray!40}{---} \\
\rowcolor{gray!5}
Next Step & Navigate forward through WE steps to observe the next step in the solution & \textcolor{gray!40}{---} & \checkmark & \textcolor{gray!40}{---} & \textcolor{gray!40}{---} \\
Prev Step & Navigate backward through WE steps to review previously demonstrated steps in the solution & \textcolor{gray!40}{---} & \checkmark & \textcolor{gray!40}{---} & \textcolor{gray!40}{---} \\
\rowcolor{gray!5}
End Problem & Complete the problem by successfully deriving the conclusion (PS), finishing example review (WE), connecting all statements (Guided), or fixing all errors (Buggy) & \checkmark & \checkmark & \checkmark & \checkmark \\
\bottomrule
\end{tabular}
\end{table}

\section{User Study}
\subsection{Experimental Design}
We conducted a controlled between-subjects study to evaluate the effectiveness of our developed interventions. We designed three training conditions in the logic tutor:

\begin{itemize}
    \item Control (\textbf{{$Control$}}): Students received PS or WE (selected randomly) during training.
    \item Treatment 1 (\textbf{{$Buggy$}}): Students received PS or Buggy (selected randomly) during training.
    \item Treatment 2 (\textbf{{$Guided$}}): Students received PS or Guided (selected randomly) during training.
    
\end{itemize}
The system recorded all student interactions with the interface, including timestamp data, step-by-step action sequences, and help-seeking behaviors.

\subsection{Participants}
The tutor was deployed with 155 students in an undergraduate Discrete Mathematics course at a public research university in the United States in the Spring of 2025. We did not collect course-specific demographics; however, Discrete Math is a mandatory course for all CS majors. Therefore, for an approximation, we report the demographics of the 2021-22 graduating class of CS majors with the gender composition of 83\% men and 17\% women; and race/ethnicity of 58\% white, 18.5\% Asian, 3\% Hispanic/Latin, 2\% Black/African American, 9\% other races, with the remaining 9.5\% having international student status for whom race/ethnicity information was not available. This study was conducted at a university where we applied for and received an exemption for conducting the study under an IRB (Internal Review Board), and only authorized researchers could access the data collected from the participants.

Each participating student was assigned to one of the three training conditions after they completed the pretest problems. We used random stratified sampling, assigning students to groups after the pretest problems. They were assigned randomly while ensuring an even distribution of students with lower and higher pretest scores across all conditions implemented that semester. We compare only students who completed all 7 levels in the tutor, with 52 students in the $Control$ group, 52 students in the $Buggy$ group, and 51 students in the $Guided$ group.

\subsection{Performance Metrics}
A student’s problem score is a combination of normalized metrics for the \textbf{problem completion time}, \textbf{solution length}, and \textbf{rule application accuracy} on a single problem, which ranks a student based on how fast, efficient, and accurate they are. This composite metric captures multiple dimensions of problem solving skills. \textbf{Problem completion time} reflects procedural fluency, the ability to recognize and apply logical rules without excessive deliberation. Faster completion times suggest automated recognition of proof patterns, a key indicator of expertise development in formal reasoning tasks. \textbf{Solution length} indicates strategic thinking and optimal problem solving skills in proof construction, providing insight into students' proof-planning abilities and their tendency toward exploratory versus expert-like problem-solving approaches. Shorter solutions indicate better strategic planning. \textbf{Rule Application Accuracy} measures conceptual understanding through the ratio of correct rule applications to total application attempts. 

The normalization ensures balanced weighting across metrics despite different scales and distributions. The resulting composite problem score helps us identify students who demonstrate balanced competency in speed, optimality, and accuracy, enabling targeted pedagogical interventions based on specific areas of weakness.

\section{Study Results}
We analyzed student \textbf{interaction logs} from three training intervention groups to measure their learning and performance.

\subsection{RQ1: Impact on Learning Outcomes}
To understand the impact of each of our training interventions on students' learning outcomes and performance, we analyzed students' score-based performance on test problems that they solved independently. Level 7 is the posttest section containing 6 test problems. To compare the performance across the three training conditions, we performed post hoc pairwise Mann-Whitney U tests \textcolor{black}{to account for the non-normal nature of our data and problem-specific variability. We reported probability-based effect sizes ($A$) for this analysis as this measure is reported to be more robust when parametric assumptions are violated \cite{ortloff2025small,haynes2021problem,coe2002s,ruscio2008probability}. This statistic represents the probability that a randomly selected participant from one group will have a higher value than a randomly selected participant from the other group. We provided $A$ values with 95\% confidence intervals ($CI$). Multiple comparisons were adjusted for using} Bonferroni’s correction $\alpha$ = 0.016\footnote{In the pairwise tests, each data point was used in at most three tests: ($Control$, $Buggy$), ($Control$, $Guided$), and ($Buggy$, $Guided$). Thus, corrected $\alpha$ = 0.05/3 = 0.016} \cite{mcknight2010mann,weisstein2004bonferroni}. \textcolor{black}{In addition to the statistical tests, we performed mixed-effects regression analysis \cite{west2022linear,harrison2018brief}. All student interactions with the tutor were recorded, thus there were no missing values in the dataset.} Note that no significant differences were found in performance across the three groups in the pretest problems.

\begin{table}[t]
\centering
\caption{Overall Problem Score and Three Metrics: Rule Accuracy, Problem Completion Time (minutes), and Solution Length (Mean (SD)) in Test Problems. [Note: \textsuperscript{**} represents $p$-value<0.0003, \textsuperscript{*} represents $p$-value<0.016, and \textsuperscript{$\dagger$} represents marginal significance (adjusted threshold after Bonferroni correction).]}
\label{tab:overall_results}
\renewcommand{\arraystretch}{1.3}
\begin{tabular}{@{}lcccc@{}}
\toprule
\textbf{Group} & \textbf{Problem Score} & \textbf{Rule Accuracy} & \textbf{Time (minutes)} & \textbf{Solution Length} \\
\textbf{(N)} & Mean (SD) & Mean (SD) & Mean (SD) & Mean (SD) \\
\midrule
\cellcolor{gray!8}$Control$ (52) & \cellcolor{gray!8}67.8 (22.2) & \cellcolor{gray!8}73.2 (22.1) & \cellcolor{gray!8}11.7 (22.6) & \cellcolor{gray!8}8.8 (3.8) \\
$Buggy$ (51) & 72.3 (21.8)\textsuperscript{*} & 76.8 (20.8)\textsuperscript{$\dagger$} & 8.4 (14.5)\textsuperscript{*} & 8.7 (3.7) \\
\cellcolor{gray!8}$Guided$ (52) & \cellcolor{gray!8}72.4 (22.4)\textsuperscript{*} & \cellcolor{gray!8}79.2 (18.6)\textsuperscript{*} & \cellcolor{gray!8}8.9 (16.1)\textsuperscript{*} & \cellcolor{gray!8}8.7 (3.5) \\
\bottomrule
\end{tabular}
\end{table}

\begin{figure}[t]
\centering
\includegraphics[width=0.8\linewidth]{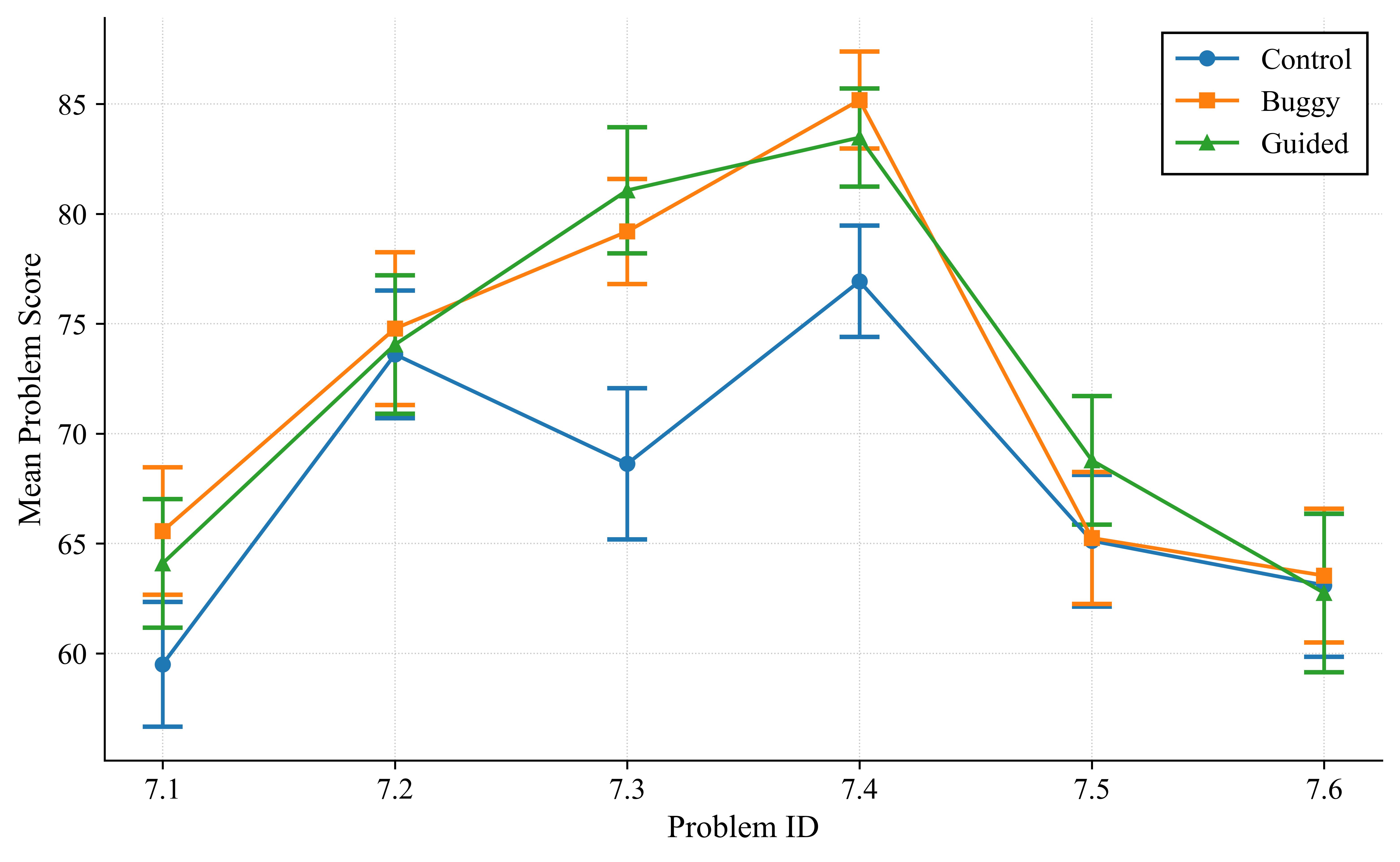}
\caption{Comparisons of Problem Score Across Conditions during Posttest Problems in Level 7 [Note: Table \ref{tab:posttest-problems} in \autoref{app:rq1} contains the same results for an alternative representation.]}\label{fig1:problemScore}
\end{figure}
\textbf{Problem Score.} Figure \ref{fig1:problemScore} shows the problem score trajectories across all three conditions. No significant differences in pretest scores indicated equivalent initial competency levels across conditions. 

\textcolor{black}{Both treatment groups showed significantly higher posttest scores compared to the Control group ($mean = 67.8$). The $Buggy$ group showed a significant improvement ($mean = 72.3$, $A = 0.56$, 95\% CI [.52, .61], $p = .009$). Similarly, the $Guided$ group also demonstrated a significant improvement ($mean = 72.4$, $A = 0.57$, 95\% CI [.52, .61], $p = .004$) (see Table \ref{tab:overall_results}).}

\textcolor{black}{In addition to the statistical tests reported, we performed a mixed-effects regression analysis \cite{west2022linear,harrison2018brief} for all posttest problems to further verify the association between the given training treatment ($Control, Buggy, Guided$) and problem scores. We used problem IDs as a random effect variable to control for the impact of differences between test problems. We used the treatment type as the fixed effect and problem score as the dependent variable. The analysis, using the $Control$ condition as the reference group, confirmed that training conditions had a significant overall effect on posttest scores. Compared to the $Control$ condition ($mean = 67.82, SE = 3.32$), participants in the $Buggy$ condition scored significantly higher ($\beta = 4.44, SE = 1.71, z = 2.60, p = .009$), as did participants in the $Guided$ condition ($\beta = 4.56, SE = 1.70, z = 2.68, p = .007$). Specifically, the unstandardized coefficient ($\beta$) for $Buggy$ indicates that participants in the $Buggy$ group scored, on average, 4.44 points higher than the $Control$ group. Similarly, the $\beta$ for $Guided$ shows an average score difference of 4.56 points compared to the $Control$ group. }


\textbf{Rule Application Accuracy.} In addition to the overall problem score, we observed students' step derivation behavior. Total correct steps are the number of correct rule applications in a problem. Total incorrect steps are the number of rule applications that are either incorrect by virtue of selecting a rule that does not apply to the arguments (e.g., using the rule \textit{Simplification} on a logical statement that cannot be simplified) or not being able to correctly identify what statement a rule application would derive. Rule application accuracy is defined as the number of total correct rule applications divided by the total number of attempts for rule application. As shown in Table \ref{tab:overall_results}, $Guided$ students achieved significantly higher rule application accuracy than $Control$ in posttest problems ($Guided$ (mean) = 79.3, $Control$ (mean) = 73.2, $p$ = 0.001\textcolor{black}{, $A = .58$, 95\% CI [.53, .62])}. The rule application accuracy was not significantly different between $Buggy$ and $Control$ in posttest problems ($Buggy$ (mean) = 76.8, $Control$ (mean) = 73.2, $p$ = 0.04\textcolor{black}{, $A = .55$, 95\% CI [.50, .60])}.

\textbf{Time.} 
$Buggy$ group had significantly lower problem completion time than the $Control$ group in posttest problems ($Control$ (mean) = 11.7 minutes, $Buggy$ (mean) = 8.4 minutes, $p$ = 0.01\textcolor{black}{, $A = .56$, 95\% CI [.51, .60])}. Similarly, $Guided$ group also had significantly lower problem completion time than the $Control$ group in posttest problems ($Control$ (mean) = 11.7 minutes, $Guided$ (mean) = 8.9 minutes, $p$ = 0.001\textcolor{black}{, $A = .56$, 95\% CI [.53, .63])}. The problem completion time in the posttest problems for the groups $Buggy$ and $Guided$ was not significantly different ($p$ = 0.24). 

\begin{figure}[ht]
\centering
\includegraphics[width=0.5\linewidth]{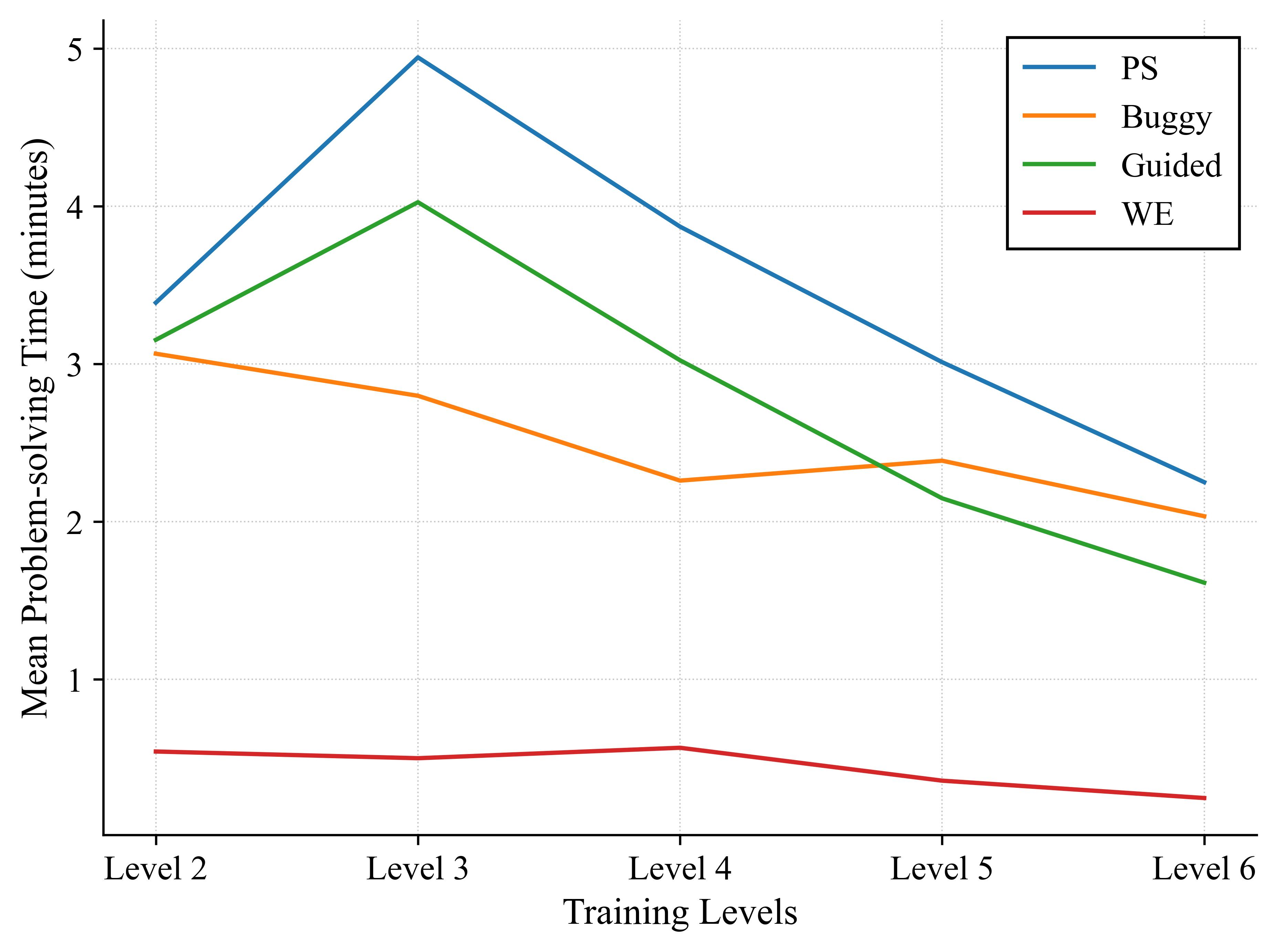}
\caption{Comparisons of Mean Problem time for Each Problem Type during Training (Level 2 through Level 6)}
\label{fig_buggy_time}
\end{figure}

As shown in Figure \ref{fig_buggy_time}, PS required the longest completion times, reflecting its open-ended nature and higher cognitive load. In contrast, WE demanded minimal time investment, as students primarily engaged in passive observation (clicking buttons to progress through the tutor steps) rather than active proof construction. The mean problem completion time for Guided and Buggy is between; students spent more time on Guided examples in the earlier phase of training compared to the mean time spent on Buggy. In the later phases of the training (level 5 and level 6), students spent less time in Guided and more time in Buggy. Both treatment groups required more training time than the $Control$ condition ($Buggy$: mean=1.12h, SD=0.5; $Guided$: mean=1.17h, SD=0.5; $Control$: mean=0.79h, SD=0.5), reflecting the additional cognitive engagement required for structured proof practice. Notably, despite spending more time during training, treatment groups achieved comparable total tutor times (around 4 hours across all conditions). More details appear in Table \ref{tab:time_three_conds} (\autoref{app:rq1}).

\textbf{Solution Length.} As shown in Table \ref{tab:overall_results}, there were no significant differences in solution lengths across groups, indicating the treatments did not help students learn to construct shorter solutions. 

\autoref{app:rq1} includes the visualization of each metric's value in posttest problems across conditions.
 
\subsection{RQ2: Interaction with Prior Knowledge}
Prior research suggests that instructional interventions like worked examples and Parsons problems may have varying effectiveness based on learners' prior knowledge \cite{kalyuga2009expertise,koedinger2012knowledge}. To investigate this phenomenon in our context, we analyzed how the impact of Buggy and Guided examples differed based on prior knowledge. We categorized two prior knowledge groups (high and low) based on a median split on the pretest score. We compared metrics across two test sections (pretest and posttest) across high and low prior knowledge groups, using pairwise post hoc Mann-Whitney U tests with Bonferroni correction (corrected $\alpha$ = 0.05/3 or 0.016). Note that a total of 18 tests (2 pretest score groups, 3 dependent variables, and 3 pairwise tests for each group and performance metric) were carried out to compare the metrics presented in Table~\ref {tab:mediation_analysis}. Thus, a more conservative Bonferroni correction could be carried out to eliminate false positives. However, we do not want to introduce many false negatives while eliminating false positives. That is why, we decided the level of Bonferroni correction based on the number of unique pairwise tests that each data point participated in (3 pairwise tests across interventions), rather than on the number of related tests \cite{bland1995multiple}. 

As shown in Table~\ref{tab:mediation_analysis}, the $Guided$ group achieved significantly improved rule accuracy for low prior knowledge students at posttest problems ($Guided$ (mean) = 78.7, $Control$ (mean) = 71.8, $p$ = 0.013\textcolor{black}{, $A = .59$, 95\% CI [.52, .65])}. High prior knowledge students showed comparable posttest rule accuracy across groups $Control$ and $Guided$ ($Guided$ (mean) = 79.7, $Control$ (mean) = 74.4, $p$ = 0.04\textcolor{black}{, $A = .56$, 95\% CI [.50, .62])}, potentially suggesting a ceiling effect for learning rules. \autoref{app:rq2} includes the visualization of rule accuracy values in posttest problems across groups. Guided examples reduced problem completion time for low prior knowledge students during posttest problems ($Control$ (mean) = 12.3 minutes, $Guided$ (mean) = 8.6 minutes, $p$ = 0.005\textcolor{black}{, $A = .60$, 95\% CI [.53, .66])}. The solution lengths were not significantly different between the $Guided$ and $Control$ groups. These results suggest that Guided examples helped novices gain a better understanding of logic rules and complete the proof in less time.

The results for the group $Buggy$ showed the opposite trend, revealing interesting aptitude-treatment interactions \cite{snow1991aptitude}. Across three dimensions (rule accuracy, problem time, solution length), there were no significant differences between groups $Buggy$ and $Control$ for the low prior knowledge students. In contrast, the high prior knowledge students in group $Buggy$ achieved significantly higher rule application accuracy than their counterparts in the $Control$ group in the posttest problems ($Buggy$ (mean) = 82.6, $Control$ (mean) = 74.4, $p$ = 0.0004\textcolor{black}{, $A = .62$, 95\% CI [.55, .68])}. High prior knowledge students in $Buggy$ group achieved significant time reductions on posttest problems ($Control$ (mean) = 11.2 minutes, $Buggy$ (mean) = 6.3 minutes, $p$ = 0.011\textcolor{black}{, $A = .58$, 95\% CI [.52, .65])}. There was no significant difference in solution length for the high prior knowledge students between the two groups ($Control$ (mean) = 8.8, $Buggy$ (mean) = 8.2, $p$ = 0.16\textcolor{black}{, $A = .55$, 95\% CI [.48, .61])}. These results suggest that Buggy examples helped the advanced students further improve their understanding of logic rules and also complete their test problems in significantly less time.

\begin{table}[H]
\centering
\small
\caption{Performance Metrics (Mean (SD)) Across Three Training Conditions, Categorized on Pretest Scores. [Note: \textsuperscript{**} represents $p$-value<0.0003, \textsuperscript{*} represents $p$-value<0.016, and \textsuperscript{$\dagger$} represents marginal significance (adjusted threshold after Bonferroni correction).]}
\label{tab:mediation_analysis}
\begin{tabular}{@{}llcccccc@{}}
\toprule
\textbf{Metric} & \textbf{Test} &
\multicolumn{3}{c}{\textbf{High Prior Knowledge}} &
\multicolumn{3}{c}{\textbf{Low Prior Knowledge}} \\
\cmidrule(lr){3-5} \cmidrule(lr){6-8}
& &
\armhdr{$Control$}{}{26} &
\cellcolor[gray]{0.90}\armhdr{$Buggy$ }{}{26} &
\cellcolor[gray]{0.90}\armhdr{$Guided$}{}{26} &
\armhdr{$Control$}{}{26} &
\cellcolor[gray]{0.90}\armhdr{$Buggy$}{}{25} &
\cellcolor[gray]{0.90}\armhdr{$Guided$}{}{26} \\
\midrule

\multirow{3}{*}{Rule Accuracy}
& Pretest   & 64.1 (31.3) & \cellcolor[gray]{0.90}61.4 (30.5) & \cellcolor[gray]{0.90}63.5 (31.2)
                        & 44.2 (29.1) & \cellcolor[gray]{0.9}39.7 (22.5) & \cellcolor[gray]{0.90}45.1 (27.7) \\
& Posttest  & 74.4 (21.7) & \cellcolor[gray]{0.90}\textbf{{82.6 (18.5)\textsuperscript{*}}} & \cellcolor[gray]{0.90} 79.7 (18.1)\textsuperscript{$\dagger$}
                        & 71.8 (22.4) & \cellcolor[gray]{0.90}71.1 (21.5) & \cellcolor[gray]{0.90}\textbf{{78.7 (19.4)\textsuperscript{*}}} \\
\midrule

\multirow{3}{*}{Problem Time (minutes)}
& Pretest   & 11.3 (14.3) & \cellcolor[gray]{0.90}12.1 (16.7) & \cellcolor[gray]{0.90}11.7 (18.1)
                        & 28.7 (28.3) & \cellcolor[gray]{0.90}19.2 (18.7) & \cellcolor[gray]{0.90}22.9 (21.4) \\
& Posttest  & 11.2 (23.1) & \cellcolor[gray]{0.90}\textbf{{6.3 (9.2)\textsuperscript{*}}} & \cellcolor[gray]{0.90}9.1 (14.7)
                        & 12.30 (22.2) & \cellcolor[gray]{0.90}10.4 (18.1) & \cellcolor[gray]{0.90}\textbf{{8.6 (18.1)\textsuperscript{*}}} \\
\midrule

\multirow{3}{*}{Solution Length}
& Pretest   & 4.9 (1.2) & \cellcolor[gray]{0.90}5.0 (1.5) & \cellcolor[gray]{0.90}4.9 (1.3)
                        & 5.5 (1.8) & \cellcolor[gray]{0.90}6.1 (2.2) & \cellcolor[gray]{0.90}7 (3.3) \\
& Posttest  & 8.8 (3.7) & \cellcolor[gray]{0.90}8.2 (3.5) & \cellcolor[gray]{0.90}8.4 (3.6)
                        & 8.74 (3.91) & \cellcolor[gray]{0.90}9.1 (3.9) & \cellcolor[gray]{0.90}9.1 (3.5) \\
\bottomrule
\end{tabular}

\vspace{2pt}\footnotesize
\end{table}

\textcolor{black}{In addition to the descriptive statistics presented in Table \ref{tab:mediation_analysis}, we conducted mixed-effects regression analyses to further confirm how posttest performance metrics differed across training conditions after accounting for students' prior knowledge. We modeled Treatment and Prior Knowledge as fixed effects and student ID as a random effect variable. Although we present performance metric scores for the pretest in Table \ref{tab:mediation_analysis}, no significant differences were found in performance metrics across the three groups in the pretest problems. The mixed-effects models confirmed differential effects across prior knowledge groups. For rule accuracy, low prior knowledge students in the $Guided$ condition achieved significantly higher scores ($\beta = 6.83$, $SE = 1.69$, $z = 4.0$, $p < .001$) compared to $Control$ ($mean = 71.8$), while $Buggy$ examples showed no benefit ($\beta = -0.72$, $SE = 2.30$, $z = -0.3$, $p = .75$) for low prior knowledge group. Conversely, high prior knowledge students benefited significantly from $Buggy$ examples ($\beta = 8.94$, $SE = 3.93$, $z = 2.3$, $p = .023$) compared to $Control$ ($mean = 74.4$), implying a significant aptitude-treatment interaction. For problem completion time, compared to $Control$ ($mean = 12.3$), $Guided$ examples significantly reduced time for low prior knowledge students ($\beta = -3.6$, $SE = 2.01$, $z = -2.0$, $p = .045$), while $Buggy$ examples significantly reduced time for high prior knowledge students ($\beta = -4.9$, $SE = 2.36$, $z = -2.1$, $p = .03$). For solution length, compared to $Control$ ($mean = 8.8$), $Buggy$ group had marginally shorter proofs for high prior knowledge students ($\beta = -0.94$, $SE = 0.51$, $z = -1.8$, $p = 0.06$), while no significant effects were found for the $Guided$ group.}


\begin{figure}[h]
    \centering

    \begin{minipage}[t]{0.263\textwidth}
        \centering
        \raisebox{1.6cm}{\includegraphics[width=\textwidth]{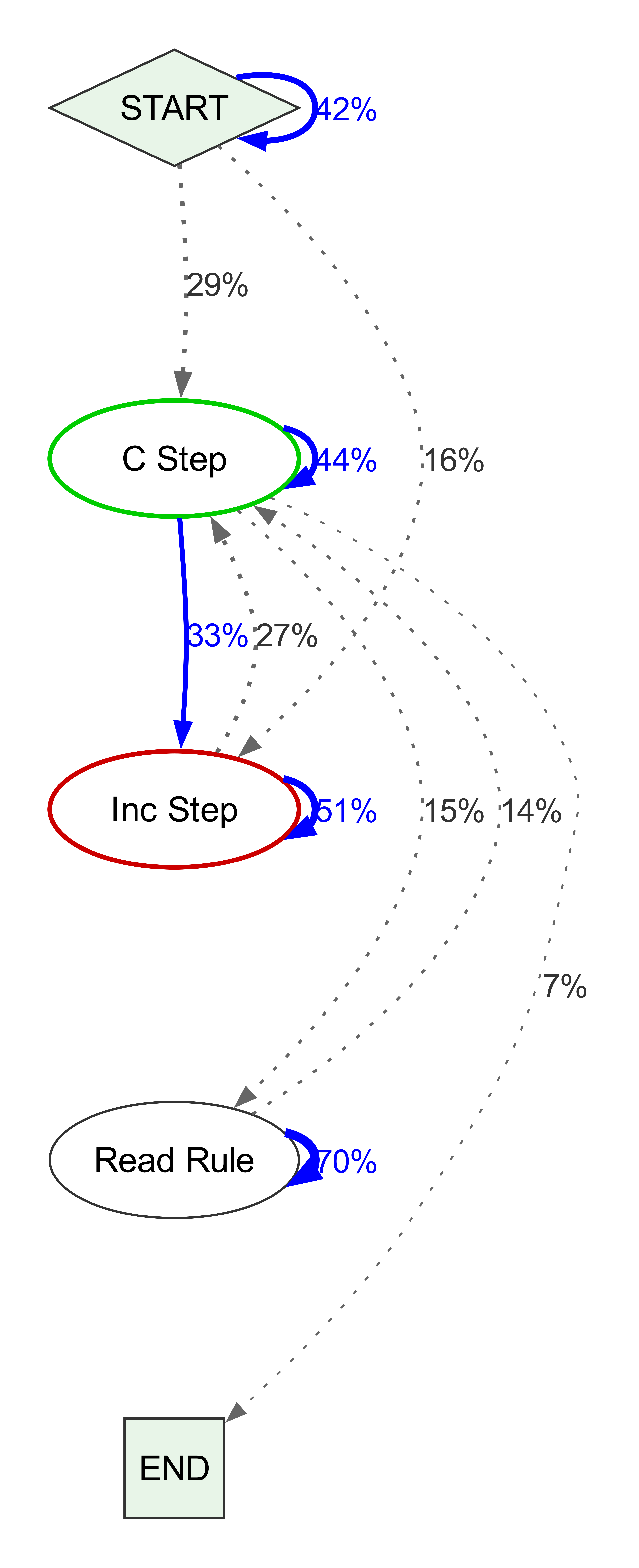}}
        \subcaption{Buggy Example (Buggy)}
        \label{fig:bpp_mini}
    \end{minipage}
    \hfill  
    \begin{minipage}[t]{0.333\textwidth}
        \centering
        \includegraphics[width=\textwidth]{Figure/guided_dot.png}
        \subcaption{Guided Example (Guided)}
        \label{fig:gpp_mini}
    \end{minipage}
    \hfill  
    \begin{minipage}[t]{0.392\textwidth}
        \centering
        \includegraphics[width=\textwidth]{Figure/PS_dot.png}
        \subcaption{Problem Solving (PS)}
        \label{fig:third_mini}
    \end{minipage}
    
    \caption{Transition Diagrams of Student Actions Across Different Problem Types}
    \label{fig:markov_mini}
\end{figure}

\subsection {RQ3: Problem Solving Behavior Across Problem Types}
The results from students' performance analysis showed that our new interventions with Guided and Buggy examples have the potential to improve students' learning performance. However, since they are a new type of problem-based training intervention, we acknowledged the necessity of analyzing their difficulty level compared to traditional training interventions such as PS or WE. Since intelligent tutoring systems are often used by students in the absence of a human tutor, we investigated further the training difficulty in different problem types, aiming to avoid any increased difficulty so that the students can persist and learn. We conducted a comparative analysis and constructed first-order Markov models from student action sequences to navigate these problems \cite{rabiner2002tutorial}. The pattern of their actions provides insights into cognitive processes that traditional performance metrics may not capture. We generated all graph visualizations with Graphviz \cite{gansner2000open} and presented them in Figure \ref{fig:markov_mini}. 

We modeled student interactions as state transitions, computing ~$P(next\ action\ |\ current\ action)$ across three problem types: PS, Buggy, Guided. We do not present the analysis for WE problems, given that students could only observe the proof steps performed by the tutor. Figure \ref{fig:markov_mini} visualizes these transition diagrams. We displayed edges with >15\% probability (preserving isolated nodes' strongest connections for graph connectivity) to balance detail with interpretability. We highlighted the transitions with probability > 30\% in blue while fading out the rest. In the diagrams, ``C Step'' represents a correct attempt, and ``Inc Step'' represents an incorrect attempt in relevant problem types.

As shown in Figure \ref{fig:bpp_mini}, Buggy examples demonstrate a distinctive "struggle pattern" characterized by tight loops on incorrect attempts and reading rule descriptions. Students frequently start with both correct attempts and incorrect attempts to fix any solution element, but they tend to remain stuck on incorrect attempts 51\% of the time, while transitioning to correct attempts 27\% of the time, and occasionally transitioning to reading rules. Similarly to incorrect attempts, once they start consulting the rules, they tend to read through the rules 70\% of the time. 

In contrast, Guided examples highlight the reliance on the subgoal hints. Incorrect attempts in Guided examples rarely lead to extended error cycles; students who make mistakes typically request hints (44\% of the time) rather than persist with more incorrect attempts. Guided scaffolding helped redirect the struggling students toward help-seeking and correct attempts. However, the strong transitions to hints from all other actions (Start, Correct Step, Incorrect Step, and Read Rule) raise the question of over-reliance on the hints in Guided examples.

Problem Solving (PS) demonstrates a more diverse set of transition patterns. A prominent pattern in PS is the frequent Delete action, with 29\% self-loops. The cycle (Correct Step, Delete, Hint, Read Rule, Correct Step) indicates frequent backtracking and revision behavior utilizing the hint system. Notably, other problem types are presented as a worked solution, so those problem representations do not support deleting given statements.

\textbf{Transition patterns in Low vs.\ High prior knowledge groups.} For each student, we generated transition probability values in each problem representation. We \textcolor{black}{conducted an exploratory analysis to investigate} the differences in transition patterns between students with low versus high prior knowledge in different problem representations. \textcolor{black}{Due to the non-normal distribution of the data, we used Mann-Whitney U tests. Given the large number of possible transitions and the exploratory nature of this analysis, we do not apply Bonferroni-adjusted thresholds}. In Table \ref{tab:student-level-diff}, we present the transition probabilities that differed between low and high prior knowledge groups with a $p$-value < 0.05 in the between-group Mann-Whitney U test \textcolor{black}{and interpret them as suggestive patterns worthy of future investigation.}


\begin{table*}[t]
\centering
\small
\renewcommand{\arraystretch}{1.2}
\caption{Transition Patterns in Low vs.\ High prior knowledge students Across Problem Types. The reported Low/High Prior Knowledge \% values are $100\times P(\text{next\ action}\mid\text{current\ action})$. Effect size $A$ denotes the probability that a randomly
selected participant from low group will have a higher value than a randomly selected participant from the high group. $A$ is asymmetric ($A_{\text{low,high}} = 1 - A_{\text{high,low}}$).}


\label{tab:student-level-diff}
\begin{tabular}{@{}l l r r r r@{}}
\toprule

\textbf{Problem Type} & \textbf{Transition} & \textbf{Low Prior (\%)} & \textbf{High Prior (\%)} & \textbf{$p$-value} & \textbf{\textcolor{black}{A, 95\% CI}}\\
\midrule

\multirow{3}{*}{PS}
 & Hint $\rightarrow$ Inc Step & 30.4 & 10.1 & .009 & \textcolor{black}{.91, [.73, 1.0]} \\
 & C Step $\rightarrow$ Delete      & 26.2 & 21.8 & .04 & \textcolor{black}{.58, [.50, .66]} \\
 & Delete $\rightarrow$ Delete      & 37.4 & 43.7 & .06 & \textcolor{black}{.40, [.30, .50]}  \\
 & Start $\rightarrow$ Hint      & 67.6 & 50.1 & .06 & \textcolor{black}{.66, [.49, .81]}\\
\midrule

\multirow{3}{*}{Buggy}
 & C Step $\rightarrow$ C Step     & 44.2 & 51.2 & .001 & \textcolor{black}{.40, [.34, .46]}\\
& Read Rule $\rightarrow$ C Step     & 32.6 & 41.8 & .02 & \textcolor{black}{.41, [.33, .48]} \\
 & Start $\rightarrow$ C Step & 64.7 & 73.7 & .03 & \textcolor{black}{.42, [.35, .49]}\\
 & C Step $\rightarrow$ Inc Step & 44.4 & 39.6 & .04 & \textcolor{black}{ .57, [.50, .63]}\\
\midrule

\multirow{2}{*}{Guided}
 & Inc Step $\rightarrow$ C Step         & 36.3 & 54.9 & .04 & \textcolor{black}{.36, [.24, .49]}\\
 & Inc Step $\rightarrow$ Read Rule & 45.8 & 58.8 & .05 & \textcolor{black}{.38, [.27, .50]}\\
\bottomrule
\end{tabular}
\end{table*}

In the PS problems, low prior knowledge students were three times more likely to transition from hints to incorrect steps than high prior knowledge students (30.4\% vs. 10.1\%, $p = 0.009$\textcolor{black}{, $A = .91$, 95\% CI [.73, 1.0]}). They also showed a stronger tendency to request hints at problem start (67.6\% vs. 50.1\%, $p = 0.06$\textcolor{black}{, $A = .66$, 95\% CI [.49, .81]}), suggesting their uncertainty about how to start the proof at the beginning. Both groups demonstrated deletion patterns, though high prior knowledge students showed slightly higher multiple deletions (43.7\% vs. 37.4\%, $p = 0.06$\textcolor{black}{, $A = .60$, 95\% CI [.50, .70]}).

In the Buggy examples, high prior knowledge students demonstrated significantly higher rates of consecutive correct steps (51.2\% vs. 44.2\%, $p = 0.001$\textcolor{black}{, $A = .60$, 95\% CI [.54, .66]}) and more successful transitions from reading rules to correct actions (41.8\% vs. 32.6\%, $p = 0.02$\textcolor{black}{, $A = .59$, 95\% CI [.52, .67]}). They were also more likely to begin problems with correct steps (73.7\% vs. 64.7\%, $p = 0.03$\textcolor{black}{, $A = .58$, 95\% CI [.51, .65]}), while low-knowledge students showed higher rates of transitioning to errors (44.4\% vs. 39.6\% for correct to incorrect transitions, $p = 0.04$\textcolor{black}{, $A = .57$, 95\% CI [.50, .63]}).

In the Guided examples, high prior knowledge students showed better error recovery, with 54.9\% of incorrect steps followed by correct ones compared to only 36.3\% for low prior knowledge students ($p = 0.04$\textcolor{black}{, $A = .64$,  95\% CI [.51, .76]}). They also more frequently read rules after errors (58.8\% vs. 45.8\%, $p = 0.05$\textcolor{black}{, $A = .72$,  95\% CI [.50, .73]}), suggesting more strategic help-seeking behavior. These patterns indicate that prior knowledge may also influence the strategic use of available support systems across different instructional interventions.

\section{Discussion}
\subsection{Main Findings}
This paper examined two different worked examples (Buggy, Guided) for learning propositional logic problem solving in an intelligent tutoring system. The results demonstrate that the choice between scaffolded practice (Guided) and debugging (Buggy) approaches has implications for both learning outcomes and problem solving strategies, with effects strongly moderated by students' prior knowledge.

\textbf{Buggy examples benefit advanced learners while Guided examples support novices.}
The interventions with Buggy and Guided examples had differential effectiveness across proficiency levels. High prior knowledge students benefited more from Buggy examples, achieving 82.6\% rule accuracy compared to 74.4\% for the $Control$ group. One factor behind this differential benefit can be the higher demand for cognitive engagement and critical thinking in Buggy examples, as a result of the limited tutor help apart from the given partial solution. In Buggy examples, students must simultaneously hold correct logical principles in mind while identifying and correcting bugs. This metacognitive reasoning exercise—distinguishing correct from incorrect—may strengthen the conceptual understanding for those who already have high prior knowledge. Research on desirable difficulties has also suggested that students can gain long-term learning benefits from more challenging materials, as long as they have sufficient knowledge or support to overcome the challenge \cite{bjork2020desirable}. Our \textit{Buggy} intervention proved ineffective for low prior knowledge students, who showed no improvement over $Control$. Without stable mental models of correct reasoning, these students may struggle to recognize bugs or may even reinforce misconceptions through exposure to Buggy examples. 

In contrast, Guided examples demonstrated more universal benefits, particularly for low prior knowledge students who achieved 78.7\% rule application accuracy in posttest problems. In Guided examples, the tutor presents a partial solution and offers step-by-step hints to complete the missing parts. Thus, the scaffolded nature of Guided examples, while requiring active participation in completing missing parts in a given solution, creates an optimal challenge level for novices. Grouping related statements into subgoal clusters may also help novices in schema construction. By offloading the construction of new statements to the tutor, students can focus their limited working memory on understanding logical relationships.

\textbf{Design recommendation.} These findings validate that learning activities in higher-order ICAP modes require higher cognitive management. \textcolor{black}{Systems should dynamically adjust interface constraints based on inferred expertise: providing structured scaffolding to novices to manage cognitive load, while introducing "productive friction" (like debugging tasks) to more proficient or expert-like users to prevent disengagement. While previous research has used AI to select activities adaptively \cite{bassen2020reinforcement,pardos2023oatutor,alam2024much} and to track students' knowledge states \cite{yudelson2013individualized,chi2011empirically,islam2025generalized,islam2024generalized}, we recommend that designers leverage real-time granular interaction logs to  predict the users' current knowledge state and cognitive requirements. For instance, systems could detect when a user is overwhelmed (requiring scaffolding) or coasting (requiring more challenge) and adapt the interaction mode accordingly.} 


\textbf{Debugging requires productive struggle, while guided proof reconstruction leads to less struggle.}
The behavioral patterns extracted from our Markov models demonstrate why these differential effects emerge. Students working on Buggy examples exhibited a productive struggle pattern of repeated attempts to identify and fix errors. Prior research demonstrated the ``productive failure'' paradigm\textemdash engaging students in solving complex problems without the provision of support structures can promote learning \cite{kapur2016examining}. \textcolor{black}{High prior knowledge students in \textit{Buggy} group showed improved scores in posttest problems, indicating that they benefited from engaging with Buggy examples.}

Guided examples, conversely, produced minimal error loops and higher help-seeking. Students who made mistakes typically requested hints (44\% probability) rather than persisting with more incorrect attempts. Research within ITSs has shown that low ability students benefit from more specific and direct guidance \cite{arroyo2000macroadapting}. The guided structure and step-specific hints in Guided examples helped the struggling students toward correct step derivation. \textcolor{black}{These results align with results from a prior controlled study showing better rule accuracy and problem solving times especially for low prior knowledge learners \cite{tithi2025investigating}. Furthermore, a thematic analysis of qualitative results in the same study showed that students felt the Parsons problems with hints simplified the problems and reduced cognitive load through guided workflows \cite{tithi2025investigating}}. This scaffolded nature may be particularly beneficial for building confidence alongside competence—a critical consideration for novice learners.

Notably, both treatment groups spent approximately 40\% more time during training than the $Control$ group yet achieved faster posttest completion times. Students who engaged with either debugging activities in Buggy examples or proof reconstruction in Guided examples developed a higher accuracy of rule application than those alternating between Problem Solving (PS) and Worked Example (WE). This finding suggests that the cognitive effort required by the interventions, while initially costly, yields superior transfer performance.

\textbf{Design recommendation.} \textcolor{black}{
Research in complex search tasks has demonstrated that analyzing \textit{nonlinear state transition patterns} allows systems to predict implicit task states, effectively distinguishing between productive exploration and struggle, without relying solely on final outcomes \cite{liu2020identifying}. Transition probabilities provide immediate, high-fidelity signals of a user’s confusion or confidence, offering a richer dataset than delayed performance scores. 
The observed struggle cycles found in our Markov analysis must be flagged in real-time as an unproductive interaction, prompting an immediate change in the instructional state. This applies not only to tutoring but also to any iterative task environment, where preventing failure loops is crucial, such as code debugging environments that could interrupt excessive "trial-and-error" attempts with contextual explanations and documentation links \cite{schoop2021umlaut}, or complex data entry workflows that prevent users from repeating the same formatting error.}


\textbf{Each problem type shapes distinct cognitive strategies through interface design.}
The distinct transition patterns across problem types reveal how interface design shapes cognitive strategies. The deletion-heavy patterns in PS\textemdash particularly for low-knowledge students\textemdash may indicate \textit{guess-and-check} strategies instead of genuine refinement strategies. Understanding these deletion patterns could inform more targeted interventions, such as evaluative feedback on deletions or metacognitive prompts about strategy selection. Our interventions\textemdash \textit{Buggy} and \textit{Guided}\textemdash each also created unique behavioral patterns: \textit{Buggy} demonstrated persistent error correction attempts, while \textit{Guided} had fewer errors and higher use of hints, indicating a potential for "gaming" behaviors or over-reliance on tutor help. 

\textbf{Design recommendation.} \textcolor{black}{
The effectiveness of the Buggy (detection/repair) and Guided (completion/structure) examples stemmed from their highly distinct interfaces, which guided students toward different cognitive goals. Interfaces should, therefore, explicitly align with their specific pedagogical or domain-specific intent. The interaction design must clearly communicate the type of thinking required, effectively scaffolding the user's mental model. For instance, this approach has been employed in broader human-AI co-creation by Suh et al., whose system \textit{Sensecape}  distinguishes between "exploration" and "sensemaking" \cite{Sensecape}. Their interface provides separate, distinct views for divergent ideation and convergent structuring, thus helping users maintain their high-level structural intent without being overwhelmed by low-level generative content.}

\subsection{Limitations and Future Directions}
Although our tutor maintains the same learning materials across different problem representations, requiring varying levels of cognitive engagement, several design limitations emerged that could be addressed in future work. 

\textcolor{black}{One significant limitation of our study is the exclusive reliance on quantitative interaction log data. While these logs allowed us to analyze specific behaviors and performance, we lack the crucial qualitative context necessary for a complete understanding of the learning process. Specifically, we do not have data regarding students' moment-to-moment affective states (e.g., frustration, confidence), their specific study environments, or detailed background characteristics beyond the clickstream data from their interaction with the tutor. This limits our ability to fully interpret the why behind the aptitude-treatment interactions—for example, whether low prior knowledge students benefited from Guided examples due to reduced confusion or increased motivation.}

The \textit{availability of scaffolding} could be more balanced across interventions; students could request hints in PS and Guided examples, but not in Buggy examples. This was a design choice to increase the cognitive challenge in Buggy examples, but the Markov model analysis revealed substantial struggle patterns when students encountered these problems. Similar to Guided examples, future iterations should introduce hints in Buggy examples when students show unproductive struggle. Moreover, prior research suggests that self-reflection and the process of \textit{making sense of errors} is potentially a more productive learning process \cite{booth2013using, grosse2007finding}. Buggy examples can be augmented with self-explanations, which may yield a better understanding of the sources of the difficulty students face in these problems.

Our results suggested that Buggy examples benefited high prior knowledge students, while Guided examples were more effective for low prior knowledge students. This finding suggests the need for \textit{adaptive problem selection} based on learner expertise. It should be noted that learners' expertise evolves throughout training; that is why a static assignment based on the pretest scores may be insufficient. Future work should implement adaptive interventions that continuously track students' knowledge growth and provide \textit{personalized} Guided examples or Buggy examples to help students improve their weak areas. \textcolor{black}{ Generative AI has shown potential to generate personalized content and hints \cite{do2025paige,park2024empowering,gu2025ai,jin2024teach,jin2025teachtune}. Integrating generative AI in tutoring systems has introduced concerns as well \cite{harvey2025don,bangerl2025creaitive,chen2025more}. Recent work evaluating LLMs for logic tutoring found that while LLMs achieved up to 86.7\% accuracy in logic proof construction, their hint explanations often lacked pedagogical appropriateness and failed to explain the high-level rationale behind hints \cite{tithi2025promise}. LLMs may generate buggy examples containing arbitrary bugs rather than pedagogically meaningful error patterns found in actual students' solutions. In Guided examples, LLMs may provide overly detailed hints that could hinder the desirable difficulty that our interventions aimed to create. Thus, leveraging generative AI in tutoring systems like ours requires a careful design of human-AI interaction and collaboration, with access to real-time student models and human-in-the-loop review, to ensure pedagogical quality. The ICAP framework can make human-AI collaborative problem solving more valuable, and help designers focus on improving interactive aspects of AI that probe and challenge human thinking to achieve the highest benefits for humans and for problem solving tasks.}

Another limitation is that, in our Markov model analysis, we only presented the transition probabilities, rather than also including the actual frequency values of each action and/or transition. Within the scope of this paper, we only presented an exploratory analysis of such fine-grained data. Future works can further investigate this data to determine ways to differentiate productive and unproductive struggle or identify scaffolding needs. Moreover, we experimented with a single tutor for a short period of time. Longer studies, such as across a semester, with multiple tutors, could generalize our findings and also reveal if the benefits from the training are transferable across domains over time.

\textcolor{black}{It is noteworthy that the observed effects in our study may change depending on factors such as differences in task type or domain complexity. Our study was conducted in a logic tutor, with problems that are considered to be open-ended but well-structured, meaning that there are many solutions but they use defined rules. We anticipate that observed benefits might be most pronounced for tasks in well-structured problem-solving domains, such as programming, math, and science. Our findings may not fully generalize to less structured domains (e.g., writing or design tasks), where the “correctness” of a step or bug is less clear. Logic proofs in our tutor usually have 12 steps. For simpler tasks, the cognitive engagement benefits of Buggy and Guided examples may diminish, as students can fix errors or complete missing connections through straightforward pattern matching rather than engaging in conceptual reasoning. On the other hand, in a highly complex problem solving domain, Buggy examples may overwhelm learners with excessive cognitive load. The expertise reversal effect we observed suggests that \textit{overall task complexity should scale with learner expertise.}}

\section{Conclusion}
In this study, we augmented a propositional logic tutor with two new types of worked examples, Buggy and Guided examples, and investigated their impacts on learning compared to passive WEs. The results show that the Buggy and Guided groups performed significantly better on the posttest problems than the Control group. Buggy problems especially helped high prior knowledge learners, whereas guided problems helped low prior knowledge learners, improving rule application accuracy and problem completion time. Through a novel application of behavior analysis, \textcolor{black}{we further investigated} student problem-solving attempts: students in the Buggy group required more persistent effort to get out of the loop of incorrect attempts. In contrast, students in the Guided group had fewer errors and higher help-seeking from the tutor. \textcolor{black}{These contrasting patterns demonstrate how different ICAP engagement modes can be strategically designed to match students' varying prior knowledge.} These findings contribute to the growing body of work on \textcolor{black}{AI-teacher and AI-student tutoring interactions} \textcolor{black}{ and provide empirical evidence that ICAP theory can guide practical design decisions in intelligent tutoring systems, and for complex domains like propositional logic where learners arrive with varying prior knowledge. Rather than simply making learning materials "more interactive," designers must consider the interaction between engagement mode and learner expertise. Designers of human-AI collaborative problem solving systems should further consider designing opportunities for challenge and assistance that can scaffold the productive struggle needed for human learning-- rather than propagating systems that promote over-reliance. These principles likely extend to} other online learning environments with high variability in users' prior experience, \textcolor{black}{and demonstrate how ICAP can be used to guide the design of complex human-AI collaborative systems for improved problem solving.}


\section{Declaration on Generative AI}
During the preparation of this work, the author(s) used ChatGPT and Grammarly to: Grammar and spelling check. After using these tool(s)/service(s), the author(s) reviewed and edited the content as needed and take(s) full responsibility for the publication's content.

\begin{acks}
Anonymized for review.
\end{acks}

\bibliographystyle{ACM-Reference-Format}
\bibliography{Main}

\appendix
\section{Appendix A}\label{app:rq1}

Table \ref{tab:posttest-problems} is added for accessibility, and contains same information as in Figure \ref{fig1:problemScore}.
\begin{table}[h]
\centering
\caption{Mean Problem Scores Posttest Problems (Level 7) Across Conditions}
\label{tab:posttest-problems}
\renewcommand{\arraystretch}{1.3}
\begin{tabular}{@{}lcccc@{}}
\toprule
\textbf{Problem ID} & \textbf{Control} & \textbf{Buggy} & \textbf{Guided} \\
& Mean (SD) & Mean (SD) & Mean (SD) \\
\midrule
\rowcolor{gray!5}
7.1 & 59.50 (2.84) & 65.57 (2.90) & 64.10 (2.93) \\
7.2 & 73.61 (2.91) & 74.78 (3.47) & 74.05 (3.15) \\
\rowcolor{gray!5}
7.3 & 68.63 (3.44) & 79.20 (2.40) & 81.07 (2.87) \\
7.4 & 76.93 (2.53) & 85.18 (2.21) & 83.47 (2.23) \\
\rowcolor{gray!5}
7.5 & 65.12 (2.99) & 65.25 (3.00) & 68.78 (2.92) \\
7.6 & 63.10 (3.25) & 63.54 (3.05) & 62.75 (3.61) \\
\bottomrule
\end{tabular}
\end{table}

\begin{figure}[ht]
\centering
\includegraphics[width=0.8\linewidth]{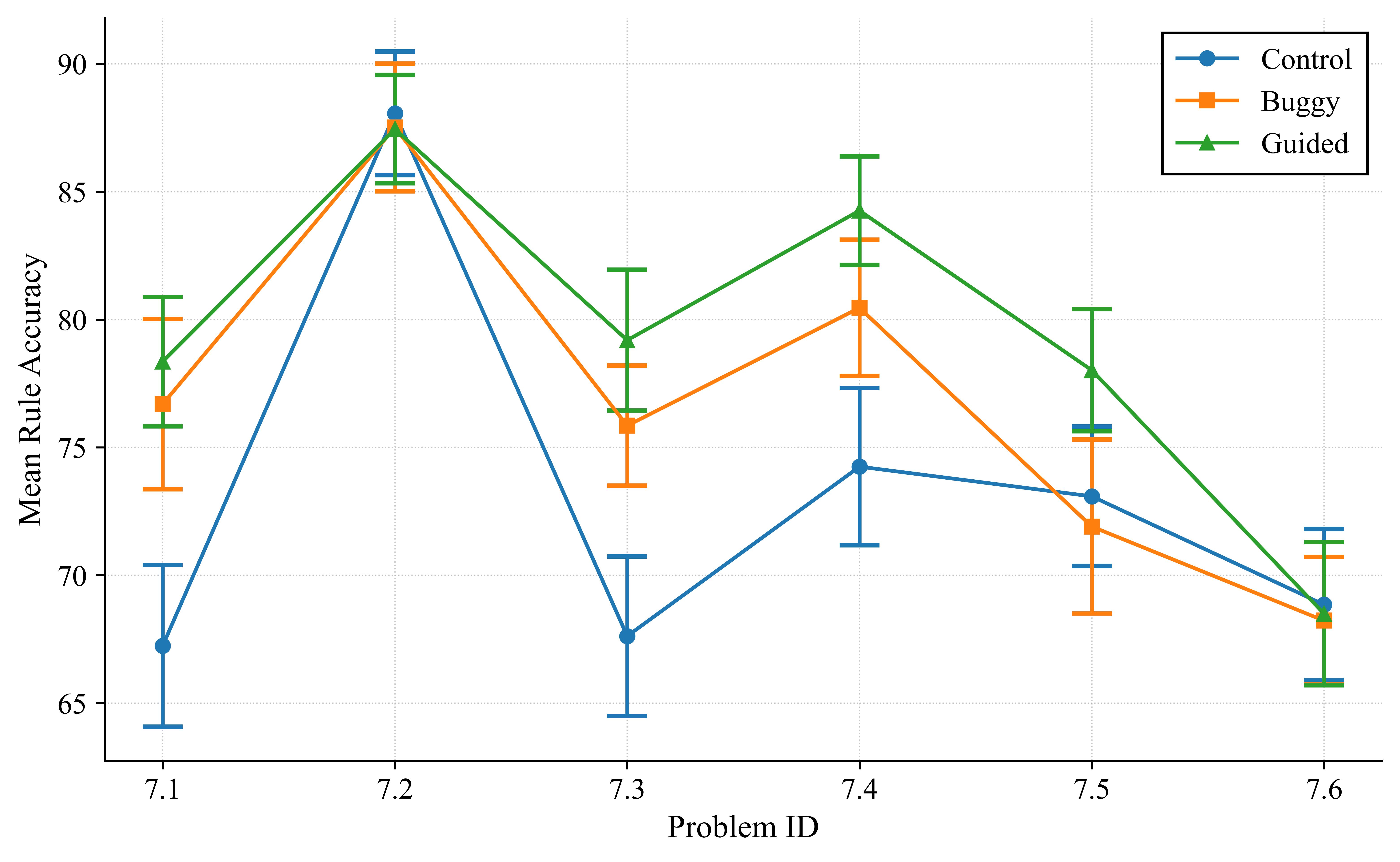}
\caption{Comparisons of Rule Accuracy Across Conditions during Posttest Problems}\label{fig1:rule_accuracy}
\end{figure}

\begin{figure}[ht]
\centering
\includegraphics[width=0.8\linewidth]{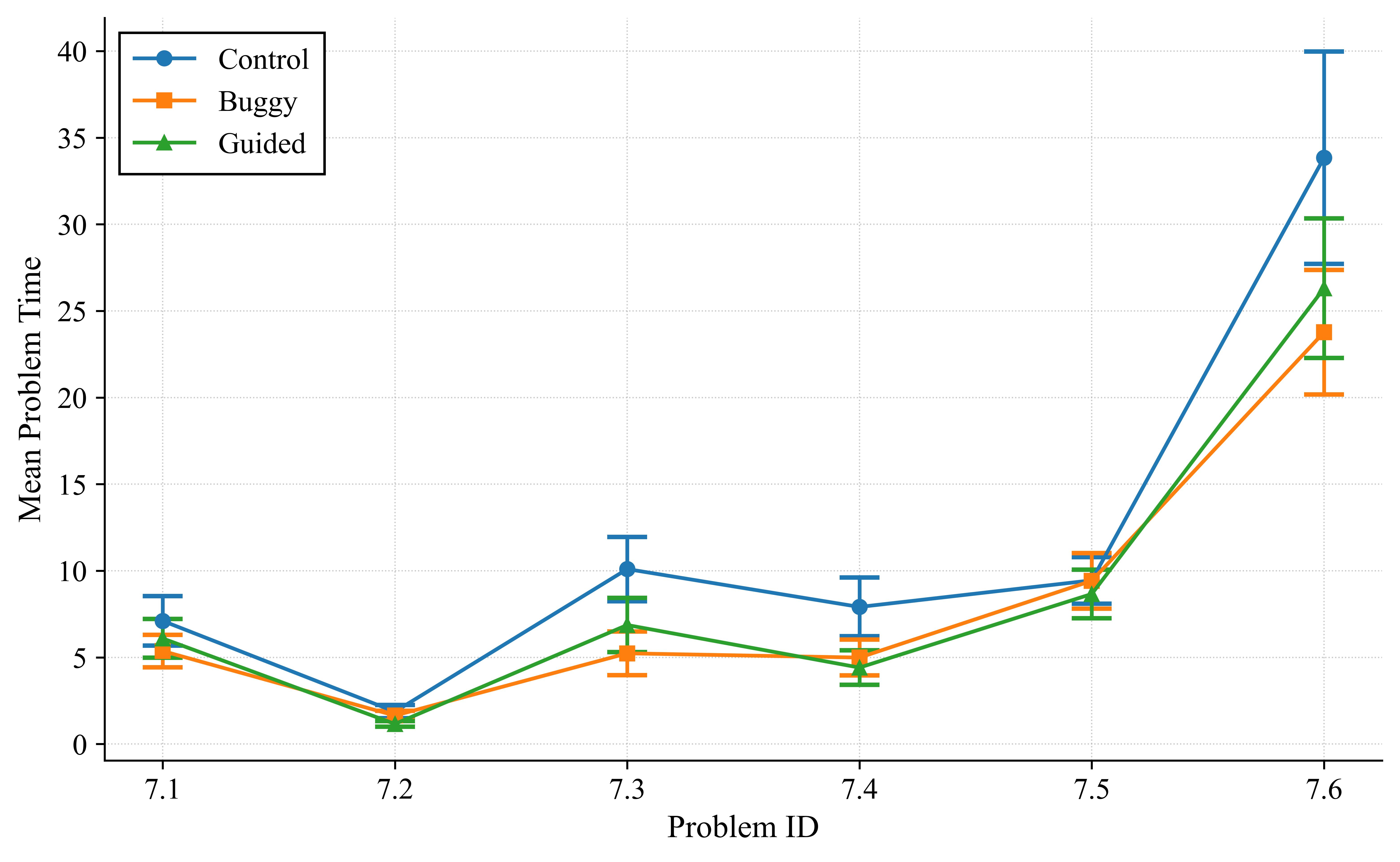}
\caption{Comparisons of Problem Time Across Conditions during Posttest Problems}\label{fig1:time}
\end{figure}

\begin{figure}[ht]
\centering
\includegraphics[width=0.8\linewidth]{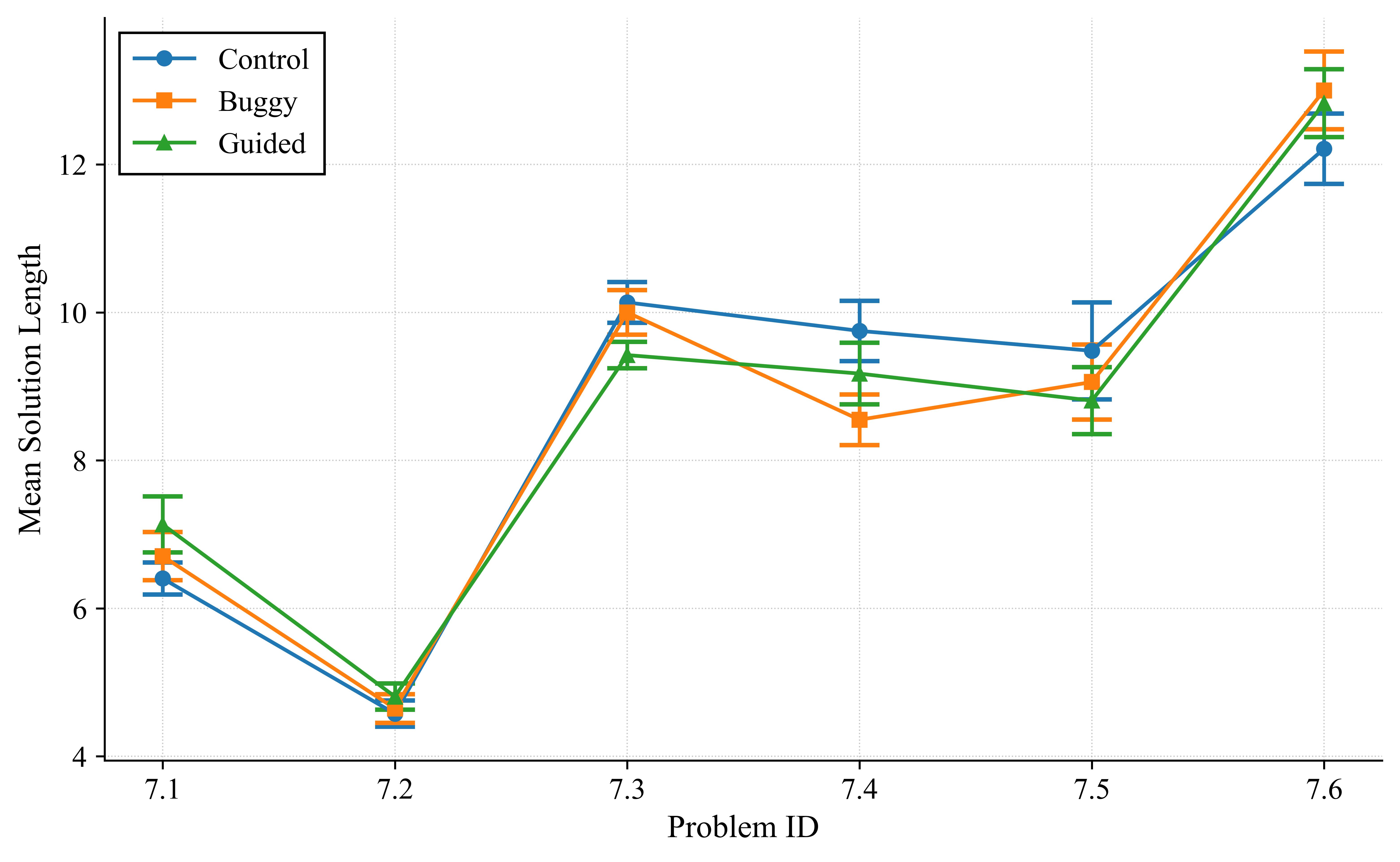}
\caption{Comparisons of Solution Length Across Conditions during Posttest Problems}\label{fig1:solution_length}
\end{figure}

Table \ref{tab:time_three_conds} presents the time spent on different sections of the tutor by each student on average.
\begin{table}[ht]
  \centering
  \caption{Time comparison across the three conditions.
           All values are hours (Mean (SD)).}
  \label{tab:time_three_conds}
  \begin{tabular}{lccc}
    \toprule
      Section & \textit{Control} & \textit{Buggy} & \textit{Guided} \\
    \midrule
      Training      & 0.79 (0.5) & 1.12 (0.5) & 1.17 (0.5) \\
      Level-End test& 1.62 (0.9) & 1.56 (0.8) & 1.43 (1.1) \\
      Posttest      & 1.17 (0.8) & 0.84 (0.5) & 0.89 (0.5) \\
      Total Tutor   & 4.24 (1.8) & 4.05 (1.43) & 4.03 (1.8) \\
    \bottomrule
  \end{tabular}
\end{table}

\section{Appendix B} \label{app:rq2}
\begin{figure}[ht]
    \centering

    \begin{subfigure}[b]{0.48\textwidth}
        \centering
        \includegraphics[width=\linewidth]{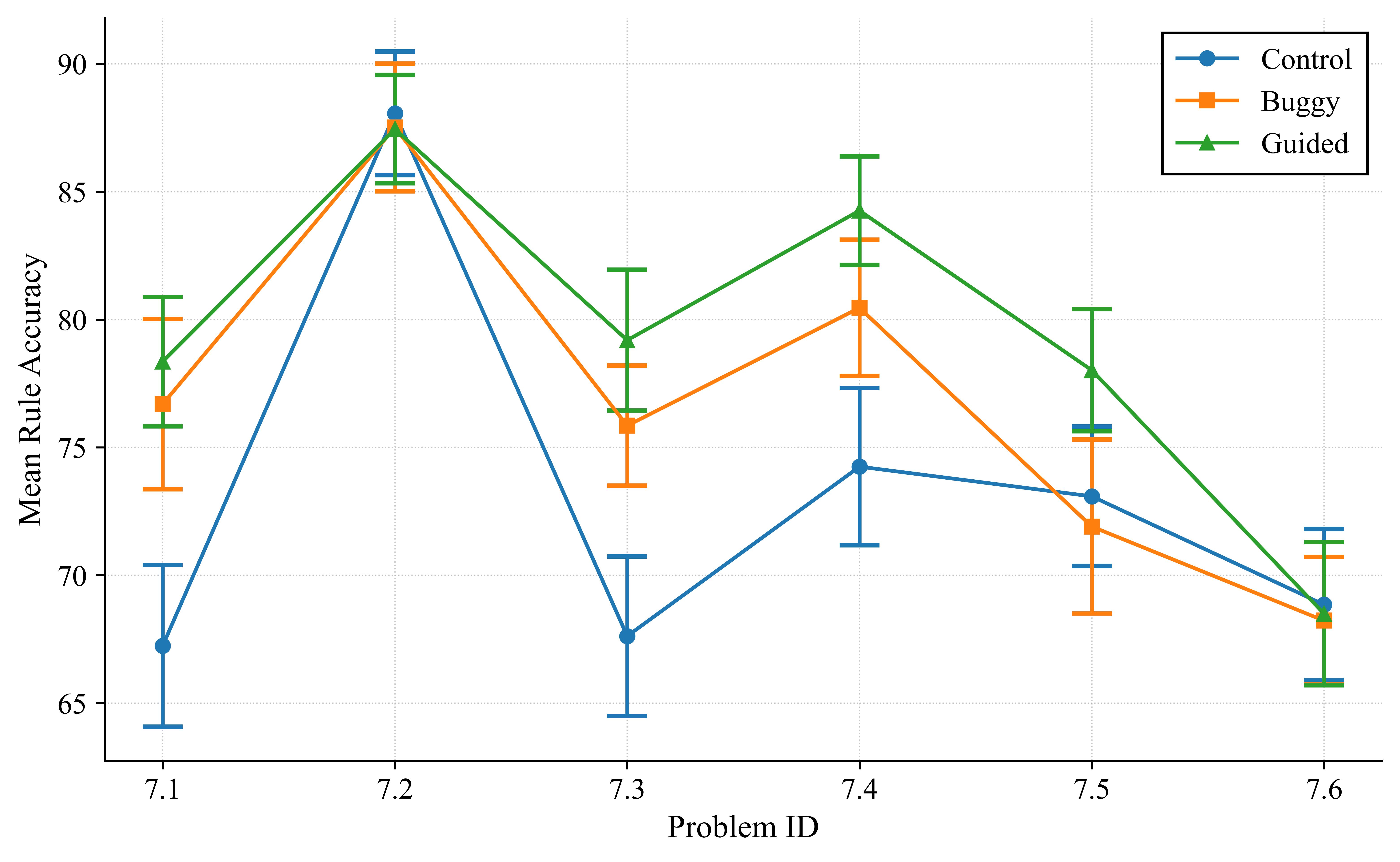}
        \caption{Upper Median}
        \label{fig:upper_ruleAcc}
    \end{subfigure}
    \hfill
    \begin{subfigure}[b]{0.48\textwidth}
        \centering
        \includegraphics[width=\linewidth]{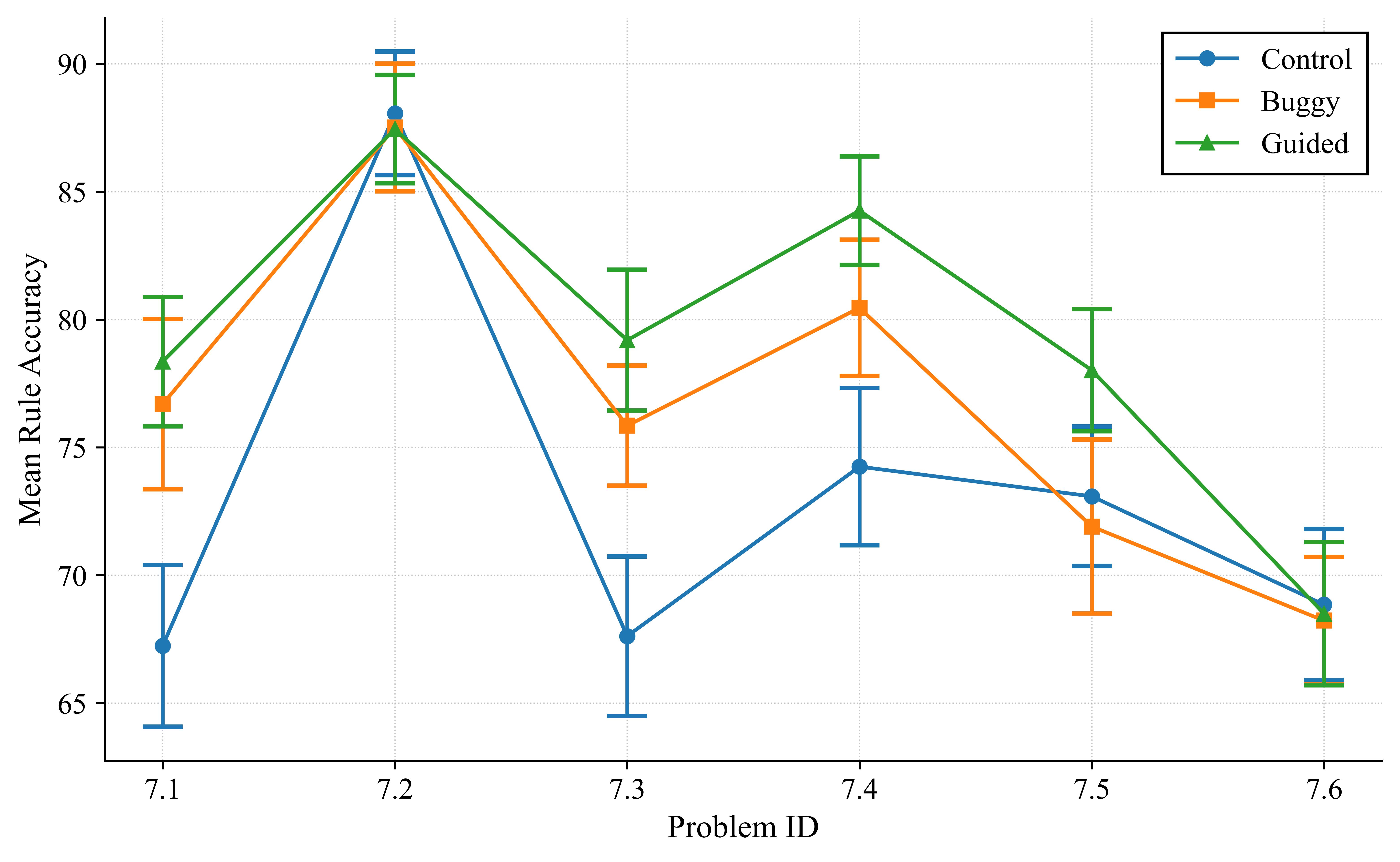}
        \caption{Lower Median}
        \label{fig:lower_ruleAcc}
    \end{subfigure}

    \caption{Comparisons of Rule Accuracy Across Conditions during Posttest Problems.}
    \label{fig:rule_accuracy_low_vs_high}
\end{figure}

\end{document}